\documentclass[%
superscriptaddress,
reprint,
amsmath,amssymb,
aps,
pre,
]{revtex4-2}

\usepackage[english]{babel}

\usepackage{graphicx}
\usepackage[percent]{overpic}
\usepackage{dcolumn}
\usepackage{bm}

\usepackage{hyperref}
\usepackage{color}
\definecolor{rltred}{rgb}{0.75,0,0}
\definecolor{rltgreen}{rgb}{0,0.6,0}
\definecolor{rltblue}{rgb}{0.3,0.3,1}

\usepackage{placeins}

\hypersetup{colorlinks,%
	hypertexnames=true,%
	linkcolor=rltred,%
	citecolor=rltgreen,%
	linktocpage=true}

\begin{document}
\title{Dynamical quasi-condensation in the weakly interacting Fermi-Hubbard model}

\author{Iva B\v rezinov\'a}
\email{iva.brezinova@tuwien.ac.at}
\affiliation{Institute for Theoretical Physics, Vienna University of Technology,
	Wiedner Hauptstra\ss e 8-10/136, 1040 Vienna, Austria, EU}

\author{Markus Stimpfle}
\affiliation{Institute for Theoretical Physics, Vienna University of Technology,
	Wiedner Hauptstra\ss e 8-10/136, 1040 Vienna, Austria, EU}
	
\author{Stefan Donsa}
\affiliation{Institute for Theoretical Physics, Vienna University of Technology,
	Wiedner Hauptstra\ss e 8-10/136, 1040 Vienna, Austria, EU}
	
\author{Angel Rubio}
\affiliation{Max Planck Institute for the Structure and Dynamics of Matter, Hamburg, Germany, EU}
\affiliation{Center for Computational Quantum Physics (CCQ), Flatiron Institute, New York, NY, USA}

\date{\today}

\begin{abstract}
We study dynamical (quasi)-condensation in the Fermi-Hubbard model starting from a completely uncorrelated initial state of adjacent doubly occupied sites. We show that upon expansion of the system in one dimension, dynamical (quasi)-condensation occurs not only for large interactions via the condensation of doublons, but also for small interactions. The behavior of the system is distinctly different in the two parameter regimes, underlining a different mechanism at work. We address the question whether the dynamical (quasi-)condensation effect persists in the thermodynamic limit. For this purpose, we use the two-particle reduced density matrix method, which allows the extension to large system sizes, long propagation times, and two-dimensional (2D) systems. Our results indicate that the effect vanishes in the thermodynamic limit. However, especially in 2D, further investigation beyond numerically tractable system sizes calls for the use of quantum simulators, for which we show that the described effect can be investigated by probing density fluctuations.
\end{abstract}

\maketitle

\section{Introduction}\label{sec:intro}
Superconductivity is one of the most intriguing collective electronic phenomena in solid state physics. The search for an explanation of high-temperature superconductivity in correlated materials still drives a considerable amount of experimental and theoretical work. Traditionally, superconductivity has been viewed as an effect of a system at (or near) equilibrium \cite{abrikosov_quantum_1965}, but pioneering experiments (see e.g.~\cite{fausti_light-induced_2011, nicoletti_optically_2014, cremin_photoenhanced_2019, rajasekaran_probing_2018, suzuki_photoinduced_2019, mitrano_possible_2016, budden_evidence_2021, buzzi_photomolecular_2020, bloch_strongly_2022}) have shown that states with signatures of superconductivity can be induced by external driving. This has stimulated new interest in dynamical condensation effects in fermionic systems (see e.g.~\cite{sentef_theory_2016,tindall_heating-induced_2019, kaneko_photoinduced_2019, cook_controllable_2020, paeckel_detecting_2020, buzzi_higgs_2021, dolgirev_periodic_2022}).\\ 
The workhorse for theoretical investigations of correlated phenomena in solid state physics is the Fermi-Hubbard model \cite{hubbard_electron_1963, hubbard_electron_1964, gutzwiller_effect_1963, kanamori_electron_1963, qin_hubbard_2022}. It has been shown analytically by Yang \cite{yang__1989} that the Fermi-Hubbard model in arbitrary dimensions has a special symmetry, the so-called $\eta$-symmetry, which gives rise to the appearance of a fermionic condensate in the excitation spectrum of the Fermi-Hubbard model. This so-called $\eta$-condensate saturates the maximally allowed occupation number for pair states, whose value was also derived by Yang \cite{yang_concept_1962}, and shows off-diagonal long range order \cite{yang__1989}. Several recent theoretical works have been devoted to showing that ground states of the Fermi-Hubbard model can be driven to metastable non-equilibrium states with large overlap with the $\eta$-condensate inheriting its properties (see e.g.~\cite{tindall_heating-induced_2019, kaneko_photoinduced_2019}).\\
Concomitantly, it was found in numerical studies \cite{rigol_emergence_2004, rodriguez_coherent_2006} and subsequently measured experimentally \cite{vidmar_dynamical_2015} that initially uncorrelated states of hard-core bosons form a quasi-condensate in one-dimension (1D) upon free expansion. This effect has been explained through the physics of emergent Hamiltonians where the dynamical system inherits the properties of a ground state condensate of the emergent non-equilibrium Hamiltonian \cite{vidmar_emergent_2017}. This study is highly relevant also for fermions since in the case of large interactions the fermions form pairs (doublons), which to a high degree of accuracy can be described as hard-core bosons. This mapping has been used to study dynamical quasi-condensation of fermions in 1D at large interactions in e.g.~\cite{heidrich-meisner_ground-state_2008,cook_controllable_2020}.\\
In this work we investigate dynamical (quasi)-condensation upon a quench with similar initial conditions, but in the opposite limit of weak electron interactions $U \ll 1 J$, where $J$ is the hopping matrix element. We show that this dynamical (quasi-)condensation effect exhibits a markedly different behavior than its counterpart at large $U$. We exploit our newly developed time-dependent two-particle reduced density matrix (TD2RDM) method \cite{lackner_propagating_2015, lackner_high-harmonic_2017, donsa_nonequilibrium_2023} to reach the required long expansion times for systems sizes of several tens of sites, while laying the ground based on small systems and exact calculations. We compare the results within the TD2RDM method with exact results for small systems to extrapolate its accuracy for larger systems where exact results are not available. Furthermore, the TD2RDM method allows to extend the investigations to two-dimensional (2D) systems enabling us to address the question whether the Mermin-Wagner-Hohenberg theorem \cite{mermin_absence_1966,hohenberg_existence_1967}, which prohibits any condensation in 1D in equilibrium in the thermodynamic limit, is also valid in this dynamical situation. \\
The paper is structured as follows: We introduce the system under investigation in Sec.~\ref{sec:sys} and review the essential building blocks of the TD2RDM method in Sec.~\ref{sec:td2rdm} emphasizing necessary extensions to incorporate the $\eta$-symmetry. We investigate dynamical (quasi-)condensation in 1D in Sec.~\ref{sec:cond_1d}. We first discuss small systems that are treatable exactly. These small systems show already signatures of the effect and serve as benchmarks for the TD2RDM method. Extensions to substantially larger systems are performed within the TD2RDM method. In Sec.~\ref{subsec:2d} we extend our research to 2D and conclude in Sec.~\ref{sec:concld}. As units we use $\hbar = m = e = 1$ unless otherwise stated.
\section{Out-of-equilibrium Fermi-Hubbard model}\label{sec:sys}
\begin{figure}[t]
	\includegraphics[width=\columnwidth]{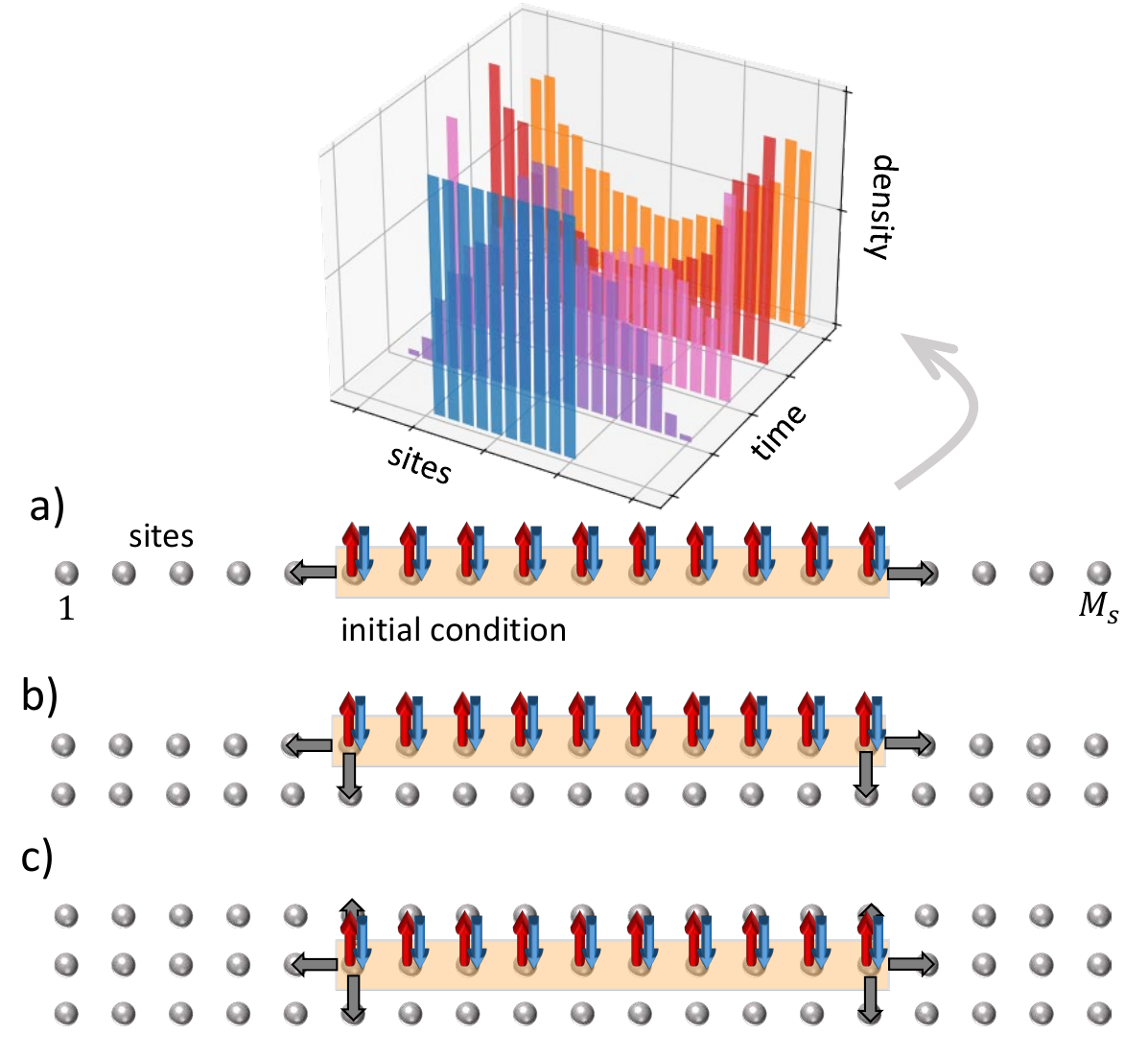}
	\caption{Fermi-Hubbard model in (a) 1D, and narrow strips in 2D with (b) two or (c) three transversal sites. The total number of sites is $M_s$. The orange bar represents the investigated initial condition of a block of adjacent doubly occupied sites either placed in the center (as shown in the images) or left aligned (not shown). The upper panel shows exemplary a few snapshots of the ensuing density fluctuations.}
	\label{fig:overview}
\end{figure}

The system under investigation is the Fermi-Hubbard model in 1D and 2D (see Fig.~\ref{fig:overview}) given by
\begin{equation}
	\hat H = -J\sum_{\langle i,j\rangle}\sum_{\sigma} \hat a_{i\sigma}^\dagger \hat a_{j\sigma} + U \sum_i \hat n_i^{\uparrow}\hat n_i^{\downarrow},
	\label{eq:ham}
\end{equation}
where $\langle i,j\rangle$ denotes nearest-neighbor hopping on a 1D or 2D lattice, and $\hat n_i^{\uparrow (\downarrow)} = \hat a_{i\uparrow(\downarrow)}^\dagger \hat a_{i\uparrow(\downarrow)}$. 
The number of sites is $M_s$. As boundary conditions we use either periodic boundary conditions or hard-wall boundary conditions. In the case of hard-wall boundary conditions, the particle density is strictly zero at the entire boundary of the system, which extends beyond the sites plotted in Fig.~\ref{fig:overview}. For example, in 1D, where the sites are enumerated from $1$ to $M_s$, the hard-wall boundary is positioned at the sites $0$ and $M_s+1$.\\
In 1D we use half-filling, i.e.~the number of particles $N$ is given by $N = M_s$ and the number of spin-up and spin-down particles is equal. As initial states $|\Psi(0)\rangle$ we use $M_s/2$ adjacent doubly occupied sites, either centered in the middle of the system as shown in Fig.~\ref{fig:overview} (a) or aligned to the left edge. Similar uncorrelated initial conditions have been used extensively as benchmarks for approximate methods to solve the time-dependent multi-particle Schr\"odinger equation, see e.g.~\cite{akbari_challenges_2012,schlunzen_nonequilibrium_2017}.\\
To explore the effect of dimensionality, we extend the 1D system by transversal degrees of freedom as depicted in Fig.~\ref{fig:overview} (b) and (c). In this case the filling corresponds to $N = M_s/2$ in Fig.~\ref{fig:overview} (b) and $N = M_s/3$ in Fig.~\ref{fig:overview} (c), and the initial conditions used in these 2D cases are those depicted in the figure.\\
All these states are initially entirely uncorrelated, i.e.~correspond to Hartree-Fock states. The correlation, however, strongly increases as a function of time as will be discussed in Sec.~\ref{sec:cond_1d}, which is key to the emergence of the (quasi-)condensate. The initial states are highly excited, i.e.~many states in the entire excitation spectrum have a non-negligible overlap. One might think that an ensuing dynamical (quasi-)condensation effect requires a particularly large overlap with the $\eta$-condensate $|\eta\rangle$, which is an exact excited eigenstate of the Hamiltonian Eq.~\ref{eq:ham}. The overlap between our initial condition and the $\eta$-condensate is, however, small as can be calculated analytically using the construction $|\eta\rangle = \left(\hat \eta^+\right)^{N/2}|0\rangle$, where $|0\rangle$ is the vacuum state and 
\begin{equation}
	\hat \eta^+ = \sum_j (-1)^j \hat a_{j\downarrow}^\dagger \hat a_{j\uparrow}^\dagger.
	\label{eq:eta_dagger}
\end{equation}
Using simple combinatorial arguments, we obtain in 1D in the case of half-filling where $N = M_s$ that
\begin{equation}
	|\langle \eta | \Psi(0) \rangle|^2 =\binom{M_s}{M_s/2}^{-1},
 \label{eq:overlap}
\end{equation}
and remains so for $t>0$. For large $M_s$ Eq.~\ref{eq:overlap} goes like $\sim 2^{-M_s}$ and is thus exponentially suppressed. Therefore, although the dynamical (quasi-)condensate described in this paper has the same energy expectation value and similar properties as the $\eta$ condensate upon its dynamical formation, it cannot be explained by an actual proximity to it.\\
To investigate the dynamics of the system in the following, we solve the Schr\"odinger equation with the Hamiltonian Eq.~\ref{eq:ham} either by using exact diagonalization or by employing the TD2RDM method as discussed below. 
\section{TD2RDM method and $\eta$-symmetry}\label{sec:td2rdm}
Within the TD2RDM method, we avoid the propagation of the wavefunction, and thus the problem of exponential scaling altogether, and instead resort to the propagation of the two-particle reduce density matrix (2RDM). The price we pay is a partial neglect of three-particle correlations, as briefly described in the following (for more details see \cite{lackner_propagating_2015, lackner_high-harmonic_2017, donsa_nonequilibrium_2023}).\\
The 2RDM is obtained from a wavefunction $|\Psi(t)\rangle$ by tracing out $N-2$ particles as
\begin{equation}
	D_{12}(t) = N(N-1)\text{Tr}_{3\hdots N}|\Psi(t)\rangle
	\langle \Psi(t)|.
	\label{eq:2rdm}
\end{equation}
Similarly, the three-particle reduced density matrix (3RDM), required for the propagation of the equations of motion of the 2RDM, is given by
\begin{equation}
	D_{123}(t) = N(N-1)(N-2)\text{Tr}_{4\hdots N}|\Psi(t)\rangle
	\langle \Psi(t)|,
	\label{eq:3rdm}
\end{equation}
and relates to the 2RDM via $D_{12} = 1/(N-2)\text{Tr}_3D_{123}$. For later reference we also introduce the one-particle reduced density matrix (1RDM) given by 
\begin{equation}
	D_{1}(t) = 1/(N-1)\text{Tr}_2D_{12} = N\text{Tr}_{2\hdots N} |\Psi(t)\rangle \langle \Psi(t)|.
	\label{eq:1rdm}
\end{equation}
In a given single-particle basis the elements of $D_{12}(t)$ can be written as 
\begin{equation}
	D_{j_1\sigma_1j_2\sigma_2}^{i_1\sigma'_1i_2\sigma'_2} = \langle \Psi(t)|\hat a_{i_1\sigma'_1}^\dagger \hat a_{i_2\sigma'_2}^\dagger \hat a_{j_2\sigma_2}\hat a_{j_1\sigma_1}|\Psi(t)\rangle.
	\label{eq:D12_sites_bas_sig}
\end{equation}
We omit the numerical index (e.g.~$12$ for the 2RDM) whenever it is obvious from the number of indices in a specific single-particle basis which RDM we consider (as in Eq.~\ref{eq:D12_sites_bas_sig}).
In the present case, a suitable basis corresponds to the individual sites of the Fermi-Hubbard model. In this basis, the uncorrelated initial conditions considered in this paper can be easily constructed as 
\begin{equation}
	D_{i\uparrow j\downarrow}^{i \uparrow j \downarrow} = 1
	\label{eq:D12_ini}
\end{equation}
for all doubly occupied sites $i$ and $j$ and zero otherwise. In the present case of a total spin singlet case it turns out that the construction of the entire $D_{12}$ is not necessary because the spin block $D_{12}^{\uparrow\downarrow}$ contains all the information \cite{lackner_propagating_2015, donsa_nonequilibrium_2023}. The equations of motion for this spin-block within the Fermi-Hubbard model are then given by 
\begin{align}
	i\partial_t D_{j_1\uparrow j_2\downarrow}^{i_1\uparrow i_2\downarrow} &= 
	\sum_n h_{n}^{i_1} D_{j_1\uparrow j_2\downarrow}^{n\uparrow i_2\downarrow} 
	+\sum_n h_{n}^{i_2} D_{j_1\uparrow j_2\downarrow}^{i_1\uparrow n\downarrow} \nonumber \\
	&+ U \delta^{i_1,i_2}D_{j_1\uparrow j_2\downarrow}^{i_1\uparrow i_2\downarrow} \nonumber \\
	&-\sum_n h_{j_1}^n D_{n\uparrow j_2\downarrow}^{i_1\uparrow i_2\downarrow} 
	-\sum_n  h_{j_2}^n D_{j_1\uparrow n\downarrow}^{i_1\uparrow i_2\downarrow} \nonumber \\
	&- U \delta_{j_1,j_2}D_{j_1\uparrow j_2\downarrow}^{i_1\uparrow i_2\downarrow} \nonumber \\
	&+ U D_{i_1\uparrow j_2\uparrow j_1\downarrow}^{i_1\uparrow i_2\uparrow i_1\downarrow}
	+ U D_{j_1\uparrow i_2\uparrow j_2\downarrow}^{i_1\uparrow i_2\uparrow i_2\downarrow}, \nonumber \\
	&- U D_{j_1\uparrow j_2\uparrow j_1\downarrow}^{j_1\uparrow i_2\uparrow i_1\downarrow}
	- U D_{j_1\uparrow j_2\uparrow j_2\downarrow}^{i_1\uparrow j_2\uparrow i_2\downarrow}
	\label{eq:eom}
\end{align}
with
\begin{equation}
	h_j^i = -J\delta_{j}^{i+1} - J \delta_{j}^{i-1}.
\end{equation}
These equations scale like $M_s^4$ with the number of sites, which is key to the efficiency of the TD2RDM method. The closure of the equations of motion requires a reconstruction of the 3RDM via the 2RDM using an approximate stable reconstruction functional \cite{lackner_propagating_2015, lackner_high-harmonic_2017, donsa_nonequilibrium_2023}. Briefly, our reconstruction functionals 
are based on the cumulant expansion of the 3RDM \cite{mazziotti_approximate_1998}
\begin{equation}
	D_{123} = \hat A D_1D_2D_3 + \hat A \Delta_{12}D_3 + \Delta_{123},
	\label{eq:3rdm_cumu}
\end{equation}
which represents a separation into elements with different levels of particle correlations. $\hat A$ is an anti-symmetrization operator that creates only permutations that give non-equivalent terms, 
\begin{equation}
	\Delta_{12} = D_{12} - \hat A D_{1}D_{2}
	\label{eq:cumu2}
\end{equation}
is the two-particle cumulant representing two-particle correlations, and $\Delta_{123}$ is the three-particle cumulant. Using the cumulant expansion, an approximate reconstruction functional of $D_{123}$ boils down to finding physically motivated reconstruction functionals for the three-particle cumulant $\Delta_{123}$. We have shown previously that a stable and accurate propagation of Eq.~\ref{eq:eom} requires that the reconstructed 3RDM correctly contracts into the two-particle space \cite{lackner_propagating_2015}. Only then are conservation of energy and spin symmetries guaranteed at all times during time propagation. This contraction consistency \cite{lackner_propagating_2015} can be achieved employing the unitary decomposition (see e.g.~\cite{absar_one_1976,harriman_geometry_1978,mazziotti_purification_2002}), which allows to decompose a tensor into its trace-free kernel and the orthogonal component (with respect to the Hilbert-Schmidt inner product), which carries traces. The orthogonal component is an exact functional of the traces of the tensor and can thus be easily determined. We employ here the reconstruction functional by Valdemoro and coworkers \cite{colmenero_approximating_1993} and enforce construction consistency \cite{lackner_propagating_2015, donsa_nonequilibrium_2023}. Enforcing contraction consistency leads to a scaling with $M_s^5$.  Due to the linearity of the unitary decomposition, this reconstruction of the 3RDM amounts to neglecting the kernel (i.e.~the trace free component) of $\Delta_{123}$ in Eq.~\ref{eq:3rdm_cumu}. We do not use the Nakatsuji-Yasuda reconstruction \cite{yasuda_direct_1997,donsa_nonequilibrium_2023} of the three-particle cumulant here because the build up of the (quasi)-condensate leads to large two-particle cumulants as a function of time without concomitant increases in the three-particle cumulants such that the Nakatsuji-Yasuda reconstruction, while being initially more accurate, leads to overall larger reconstruction errors as time progresses.\\
The $\eta$-symmetry of the Hamiltonian leads to a further constant of motion that has to be considered, i.e.~$[\hat H,\hat \eta^+ \hat\eta^-] = [\hat H,\hat \eta_z] = 0$ with $\eta^+$ given in Eq.~\ref{eq:eta_dagger}, $\eta^-$ being its hermitian conjugate, and $\hat \eta_z = \frac{1}{2}\sum_j\left(\hat n_{j\uparrow} + \hat n_{j\downarrow}-1\right)$ \cite{yang__1989}. While the conservation of the expectation value of $\langle \eta_z\rangle$ is guaranteed by the conservation of the particle number, the conservation of $\langle \eta^+ \eta^-\rangle$ requires further scrutiny. Using the 2RDM the expectation value $\langle \eta^+ \eta^-\rangle$ can easily be calculated as
\begin{equation}
	\langle \hat \eta^+ \hat\eta^-\rangle = \sum_{i,j} (-1)^{i-j}D^{i\uparrow i\downarrow}_{j\uparrow j\downarrow},
	\label{eq:eta_consv}
\end{equation}
and its behavior within the TD2RDM can be investigated by calculating its time-derivative. The fact that the time-derivative of Eq.~\ref{eq:eta_consv} has to be zero, however, does not lead to further constraints on the 3RDM reconstruction functional as it requires only exchange symmetry. (This is in contrast to e.g.~the conservation of energy \cite{akbari_challenges_2012} which requires contraction consistency \cite{lackner_propagating_2015}). However, the TD2RDM suffers from the N-representability problem, which necessitates the application of purification \cite{mazziotti_purification_2002,lackner_high-harmonic_2017, joost_dynamically_2022, donsa_nonequilibrium_2023}. We, therefore, have to adapt our purification procedure, which so far took only into account contraction consistency between the 2RDM and the 1RDM as well as energy conservation \cite{joost_dynamically_2022, donsa_nonequilibrium_2023}. The modified purification procedure, which guarantees the conservation of all constants of motion including $\langle \eta^+ \eta^-\rangle$, is described in App.~\ref{app:eta_sym}. 
\section{Dynamical (quasi-)condensation}\label{sec:cond_1d}
Our figure of merit to asses the presence of a ferimionic (quasi-)condensate is the largest eigenvalue of the 2RDM $g_1$ (i.e.~the highest geminal occupation number) obtained from diagonalization of the 2RDM
\begin{equation}
	D_{12}(t) = \sum_{j=1}^{r(r-1)/2} g_j(t)|g_j(t)\rangle
	\langle g_j(t)|,
\end{equation}
where $|g_j(t)\rangle$ are the geminal states (pair-states) and $r$ is the number of single-particle basis states (orbitals).
In our case $r=2M_s$. In our convention, the geminal occupation numbers are ordered in descending order and $g_1$ is the largest occupation number. The geminals can be grouped according to their spin symmetry into spin-singlet and spin-tripled states. In all cases reported below the largest geminal occupation number $g_1$ belongs to a spin-singlet state $|g_1\rangle$.\\
Yang showed \cite{yang_concept_1962} that $g_j$ is bounded from above by 
\begin{equation}
	g_\text{max} = \frac{N(r-N+2)}{r}
	\label{eq:gmax_gen}
\end{equation}
for a system of $N$ fermions in $r$ modes and with the normalization of the 2RDM as given in Eq.~\ref{eq:2rdm}. Clearly, $g_\text{max}$ is macroscopic, i.e.~$O(N)$, showing that the smallest reduced density matrix allowing macroscopic occupations in fermionic systems is the 2RDM, in contrast to bosonic condensation,  which manifests itself by a macroscopic occupation of an eigenstate of the 1RDM according to Penrose and Onsager \cite{penrose_bose_1956}. The maximal geminal occupation number for bosonic systems thus scales as $O(N^2)$.\\
In case of half-filling, $N = r/2 = M_s$, we obtain 
\begin{equation}
	g_\text{max} = \frac{N+2}{2},
	\label{eq:gmax_halff}
\end{equation}
and for the other cases considered here, i.e.~$N = r/4 = M_s/2$ and $N = r/6 = M_s/3$, we obtain $g_\text{max} = (3N+2)/4$ and $g_\text{max} = (5N+2)/6$, respectively.\\
When $g_\text{max}$ is reached, all other $r(r-1)/2-1$ states have equal weight given by \cite{coleman_reduced_2000}
\begin{equation}
	\bar g = \frac{2N(N-2)}{r(r-2)}.
	\label{eq:g_equal}
\end{equation}
It has been shown that states reaching $g_\text{max}$ correspond to extreme antisymmetric geminal power (AGP) states \cite{coleman_reduced_2000}, which can be constructed by antisymmetrization of a product ansatz of one pair-state. The $\eta$-condensate is an example for the realization of these states.\\
In the search for the dynamical (quasi-)condensation we will monitor the time evolution of $g_1(t)$, whose proximity to $g_\text{max}$ will signify the presence of a fermionic (quasi-)condensate. 
\begin{figure}[t]
	\includegraphics[width=\columnwidth]{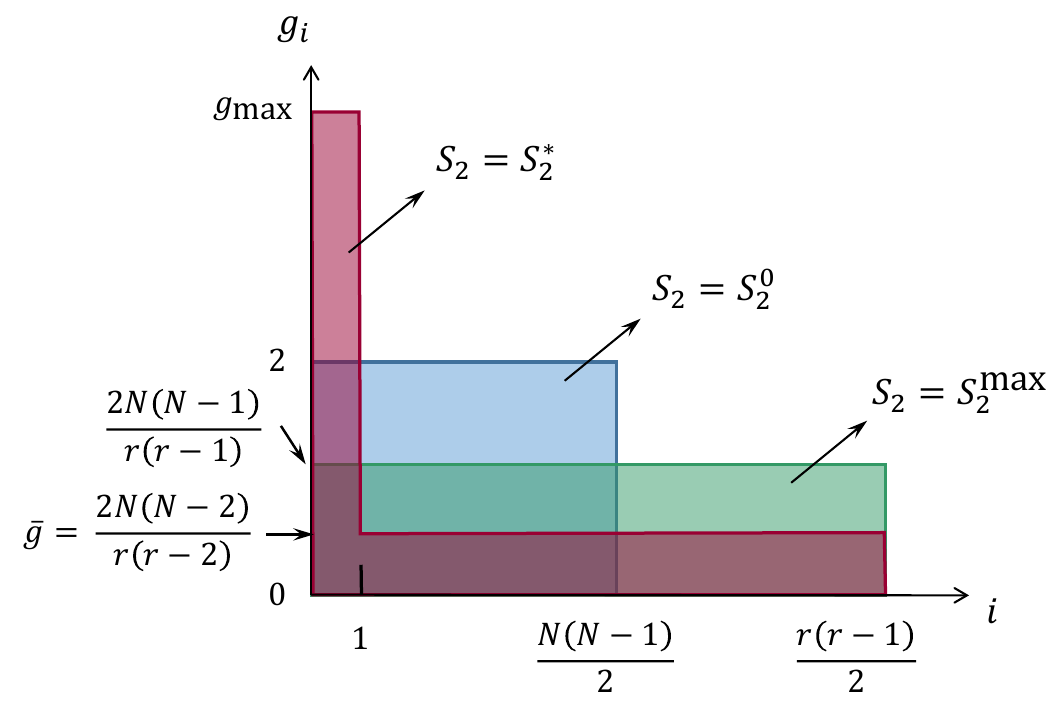}
	\caption{Distribution of geminal occupation numbers for several distinct states. Blue: uncorrelated (Hartree-Fock) state with entropy $S_2^0$, red: extreme AGP state that reaches the predicted upper bound of $g_\text{max}$ (Eq.~\ref{eq:gmax_gen}) and has an entropy of $S_2^*$, and green: a state with the highest entropy $S_2^\text{max}$ (Eq.~\ref{eq:S2}).}
	\label{fig:g_distr}
\end{figure}
When $g_1(t)$ approaches $g_\text{max}$, all other eigenvalues should become close to equal. This behavior can be analyzed by means of a single quantity, i.e.~the entropy of the distribution of geminal occupation numbers (see Fig.~\ref{fig:g_distr})
\begin{equation}
	S_2(t) = -\sum_{i=1}^{r(r-1)/2} g_i(t)\ln{g_i(t)}.
	\label{eq:S2}
\end{equation}
We evaluate $S_2$ by renormalizing $\sum_{i=1}^{r(r-1)/2}g_i = 1$. $S_2$ has been used as a measure of entanglement and correlations in the context of fermionic systems in e.g.~\cite{gigena_many-body_2021, ferreira_quantum_2022}. We denote the 2RDM entropy for an extreme AGP as $S_2^*$. Note that $S_2^*$ is close to but not equal to the maximum $S_2^\text{max}$ of $S_2$ given by an equal distribution of $g_i$, see Fig.~\ref{fig:g_distr}.
\subsection{Small systems in 1D}\label{subsec:small}
\begin{figure}[t]
	\includegraphics[width=\columnwidth]{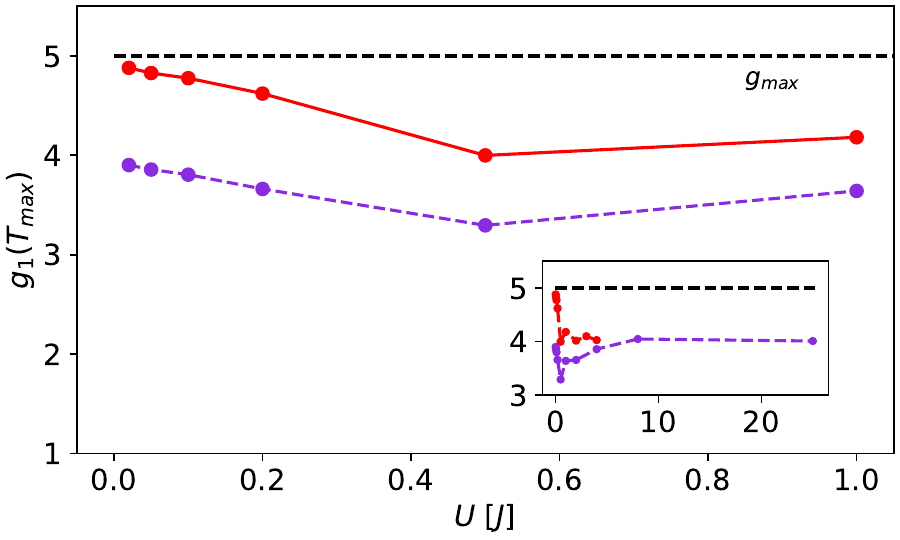}
	\caption{The maximal value of the largest occupied geminal at $T_\text{max}$, $g_1(T_\text{max})$, for different $U$ for the 1D Fermi-Hubbard model with $M_s=8$, hard-wall boundary conditions, and a left aligned initial state. Solid lines are TD2RDM results, dashed lines correspond to exact results. The black dashed horizontal line marks the maximal possible two-state occupation number $g_\text{max} = 5$ for $M_s = 8$. The inset shows the full analyzed $U$ interval (the exact results extend towards larger values).}
	\label{fig:N8_amp_Tmax}
\end{figure}

We start our investigation with $M_s = 8$ sites and exact results.
\begin{figure}[t]
	\includegraphics[width=\columnwidth]{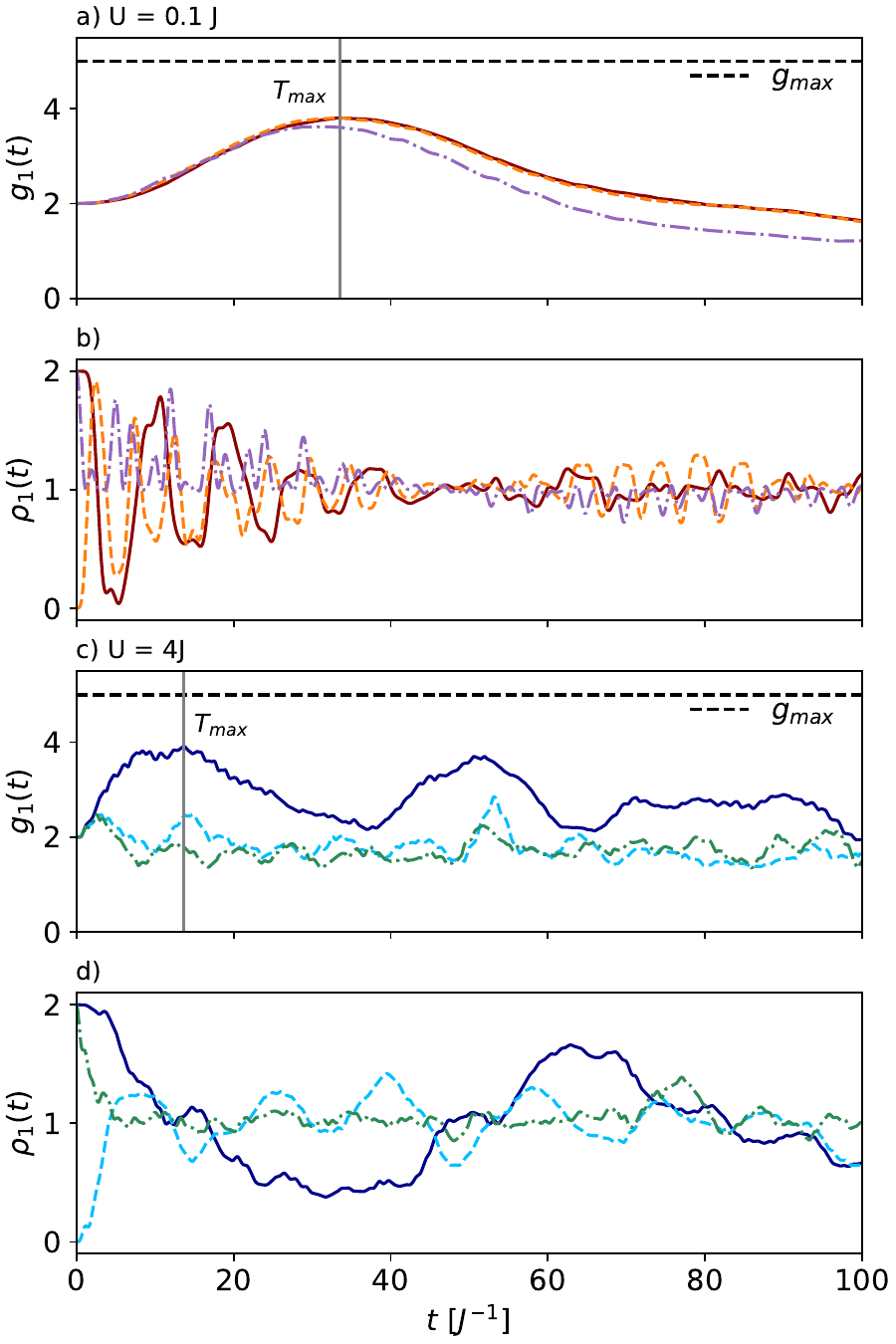}
	\caption{Exact results for the $M_s = 8$ site Fermi-Hubbard model in 1D for different $U$, boundary conditions, and alignments of the initial state. (a) and (b) $U=0.1J$, (c) and (d) $U=4J$. Plotted is  the maximal geminal occupation number as a function of time, $g_1(t)$, in (a) and (c), and the particle density at site one as a function of time, $\rho_1(t)$, in (b) and (d). In all sub-figures solid lines correspond to hard wall boundary conditions and a left aligned initial state, dashed lines correspond to hard wall boundary conditions and a centered initial state [as in Fig.~\ref{fig:overview} (a) for $M_s=8$], and dashed-dotted lines corresponds to periodic boundary conditions and left aligned initial state.}
	\label{fig:N8_maxgem_den}
\end{figure}
Tuning $U$ from $U\ll 1J$ to $U\gg 1J$ we observe significant differences in the system's behavior, see Fig.~\ref{fig:N8_maxgem_den}. For $U = 4J$ and periodic boundary conditions, $g_1(t)$ fluctuates around $g_1(t)\approx 2$ as a function of time regardless of the position of the initial state [see Fig.~\ref{fig:N8_maxgem_den} (c) for the example of a left-aligned initial state]. The system thermalizes within $t \approx 5 J^{-1}$, soon after the particle density completes a full circle around the system [Fig.~\ref{fig:N8_maxgem_den} (d)]. This behavior is also observed for hard wall boundary and a centered initial condition, with thermalization occurring before any significant increase in $g_1(t)$.\\
Significantly different behavior is observed when employing a left-aligned initial state and hard wall boundary conditions. In this scenario, more space (i.e.~half of the system) is available for the system to expand freely until reaching the boundary. During this period, a noticeable enhancement of the pair-occupation number occurs, until $g_1$ reaches at the time $T_\text{max}$ a maximum of about $g_1(T_\text{max}) \approx 4$. Here and in the following, $T_\text{max}$ represents the time of the first significant maximum in $g_1(t)$ [see Fig.~\ref{fig:N8_maxgem_den} (c)]. Overall, the process of thermalization is slower, allowing for the observation of a second local maximum during the "back-and-forth" movement of the density, compare Fig.~\ref{fig:N8_maxgem_den} (c) and (d).
These observations are in line with the physics of emergent Hamiltonians \cite{vidmar_emergent_2017}, elucidating the phenomenon of dynamical quasicondensation in the context of hard-core bosons upon free expansion \cite{rigol_emergence_2004}: The large interaction parameter $U$ induces a pairing between the fermions, facilitating a mapping to hard-core bosons, also known as doublons (see e.g.~\cite{cook_controllable_2020}). If the system can expand freely for long enough time, quasicondensation emerges. However, the interaction with the boundary ultimately destroys the validity of the emergent Hamiltonian and with it the quasicondensate.\\
In our analysis of the dynamical (quasi-)condensation effect, we focus on evaluating $T_\text{max}$, and the associated amplitude of the geminal occupation number $g_1(T_\text{max}$). For $U\geq 1 J$ we observe that $T_\text{max}$ scales linearly with $U$, i.e.~$T_\text{max}\propto U$ [Fig.~\ref{fig:N8_Tmax} (b)]. This behavior can be attributed to the rescaled hopping matrix element of the doublons, which is proportional to $J^2/U$ (see e.g.~\cite{cook_controllable_2020}). The dynamical (quasi-)condensation effect is least pronounced in the range of $U\approx 1 J$, where we observe a substantially smaller enhancement of $g_1(T_\text{max})$ to values around $g_1(T_\text{max})\gtrsim 3.5$ (Fig.~\ref{fig:N8_amp_Tmax}). As $U$ increases, $g_1(T_\text{max})$ shows a steady rise, reaching saturation at $g_1(T_\text{max})\approx 4$.

\begin{figure}[t]
	\includegraphics[width=\columnwidth]{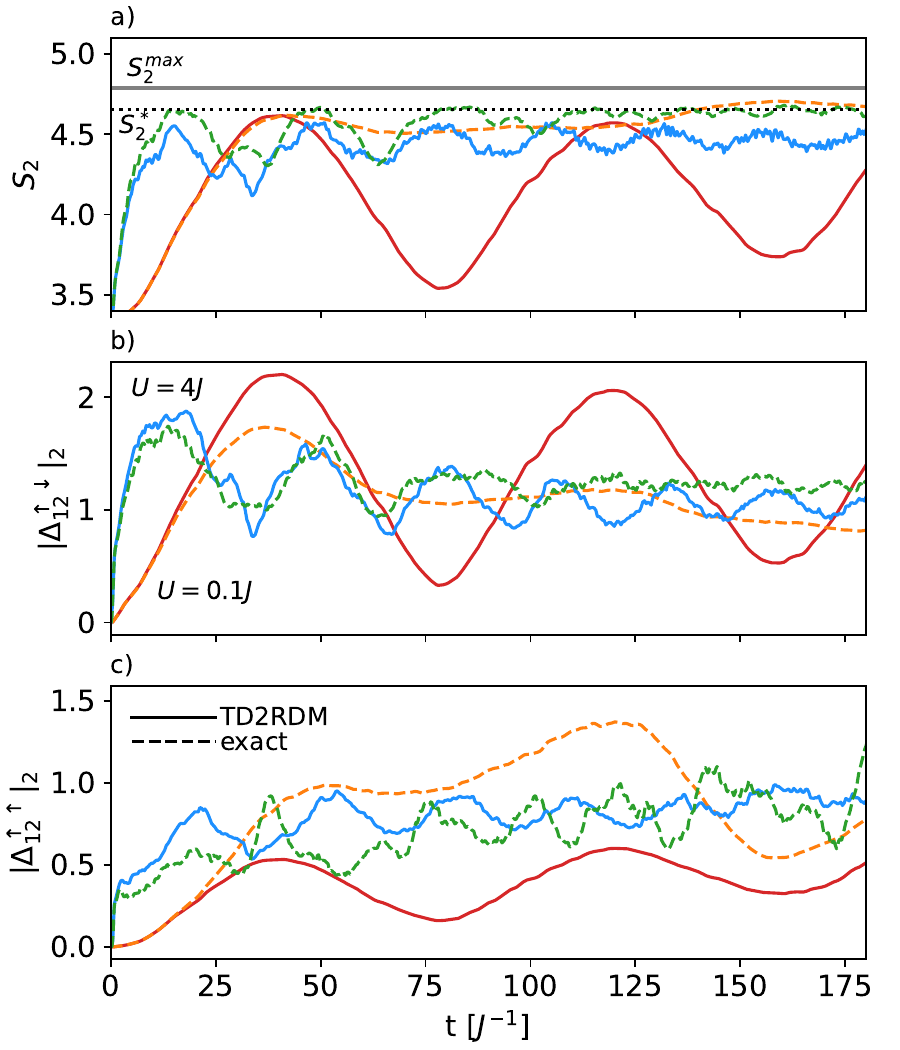}
	\caption{Comparison between exact (dashed lines) and TD2RDM results (solid lines) for (a) the dynamics of the entropy $S_2$, and the Frobenius norm of two-particle cumulants (b) $|\Delta_{12}^{\uparrow\downarrow}|_2$ and (c) $|\Delta_{12}^{\uparrow\uparrow}|_2$. In (a) the solid horizontal line marks the maximal value of $S_2$, $S_2^\text{max}$, and the dotted horizontal line denotes the value of $S_2$ obtained for a perfect condensate, $S_2^*$. Red and orange lines are for $U=0.1J$, green and blues lines are for $U=4J$. The system corresponds to the 1D Fermi-Hubbard model with $M_s=8$ sites, hard wall boundary conditions and a left aligned initial state.}
	\label{fig:N8_entropies_cumulants}
\end{figure}

For small $U< 1 J$, the behavior of the system is strikingly different. We observe the emergence of a broad maximum, with $g_1(t)$ approaching $g_\text{max}$ independently of the initial state's position and the chosen boundary conditions [Fig.~\ref{fig:N8_maxgem_den} (a)].
\begin{figure}[t]
	\includegraphics[width=\columnwidth]{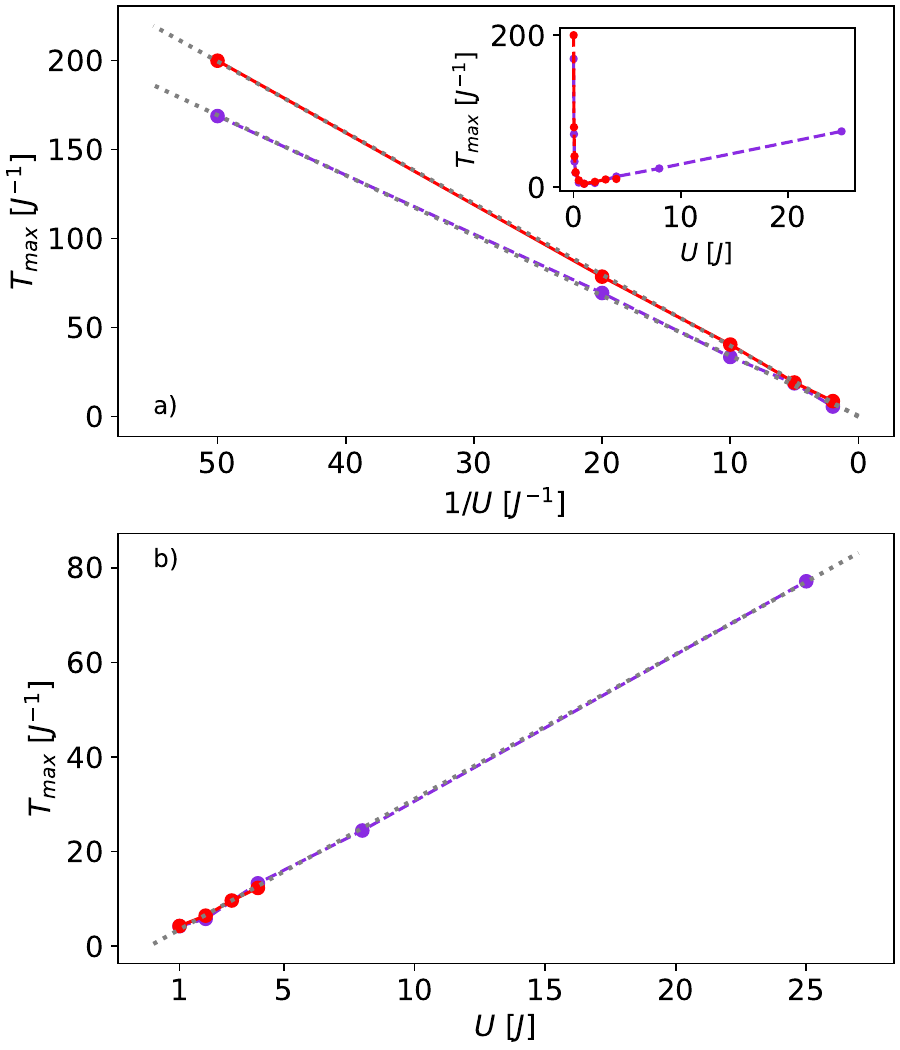}
	\caption{The time of the first maximum in $g_1(t)$, $T_\text{max}$, (see Fig.~\ref{fig:N8_maxgem_den}) as a function of $U$ for the 1D Fermi-Hubbard model with $M_s=8$, hard-wall boundary conditions, and a left aligned initial state. Solid lines are TD2RDM results, dashed lines correspond to exact results. The gray dotted lines represent linear fits. (a) Interval $U\in[0.02,0.5] J$ with inverted axis, (b) $U\in[1,25] J$. The inset in (a) shows the full $U$ interval in linear scale. The gray dotted lines in each figure correspond to linear fits.}
	\label{fig:N8_Tmax}
\end{figure}
The specific quantitative value of $T_\text{max}$ is unaffected by the initial state's position and exhibits only a mild dependence on the selected boundary conditions. Notably, the particle density undergoes multiple fluctuations between the systems boundaries during the development of the broad maximum in $g_1(t)$, without thermalizing the system. This observation leads us to the conclusion that this phenomenon represents a novel dynamical (quasi-)condensation effect, distinct from those captured by an emergent Hamiltonian. We present a detailed analysis that further supports this statement in App.~\ref{app:emer_hamil}, where we show that the Hamiltonian Eq.~\ref{eq:ham} for $U<1$ does not lead to an emergent time-depedent Hamiltonian that is (approximately) conserved during time evolution as required by \cite{vidmar_emergent_2017}.\\
Accordingly, in contrast to the linear scaling with $U$ for $U> 1 J$, we observe an inverse scaling behavior of $T_\text{max}$ for $U< 1 J$, i.e.~$T_\text{max}\propto 1/U$, see Fig.~\ref{fig:N8_Tmax} (a).
$g_1(T_\text{max})$ in turn, increases with decreasing $U$ reaching again values close to $g_1(T_\text{max}) \approx 4$.\\
We now proceed to analyze whether these observations are captured by our approximate TD2RDM method, as an underpinning for our studies with larger systems, where the effect is more pronounced. For $U> 1 J$ we observe that the TD2RDM method is in excellent agreement with the exact results, accurately predicting the linear dependence of $T_\text{max}$ on $U$ and the slope. For $U< 1 J$ the TD2RDM method correctly predicts the $1/U$ behavior but slightly overestimates the proportionality constant. Simultaneously, the amplitude at $T_\text{max}$ is overestimated but the overall behavior is again captured very well, see Fig.~\ref{fig:N8_amp_Tmax}.\\
To gain a deeper understanding of these discrepancies, we further scrutinize the dynamics of all geminal occupation numbers as predicted by the approximate TD2RDM method and compare them to the exact results.
\begin{figure}[t]
	\includegraphics[width=\columnwidth]{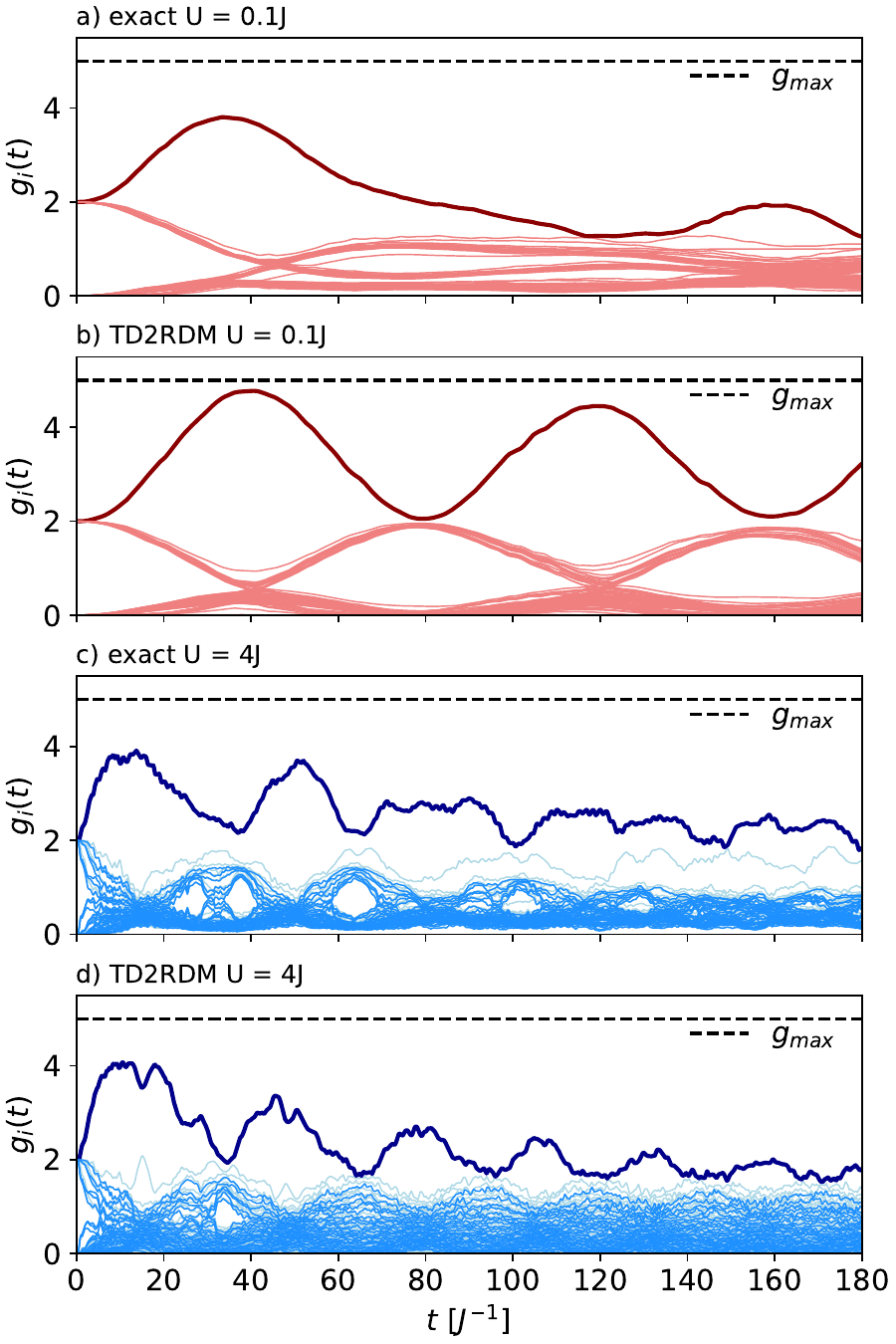}
	\caption{Comparison between exact and TD2RDM results for the dynamics of all geminal occupations numbers $g_i(t)$ for the 1D Fermi-Hubbard model with $M_s=8$, hard-wall boundary conditions and a left aligned initial state. The largest geminal occupation number $g_1(t)$ is drawn with a thicker line and darker color. (a) Exact results for $U=0.1J$, (b) TD2RDM results for $U=0.1J$, (c) exact results for $U=4J$, and (d) TD2RDM results for $U=4J$.}
	\label{fig:N8_all_geminals}
\end{figure}
We observe that as $g_1(t)$ evolves towards its maximum, all other geminal occupation numbers approach each other, consistent with the expectation that when $g_\text{max}$ is reached, all other geminals are occupied by $\bar g$, Eq.~\ref{eq:g_equal} [see Fig.~\ref{fig:N8_all_geminals} (a)]. However, the exact results show a small time shift between the maximum of $g_1(t)$ and the time at which all other $g_{i>1}(t)$ are closest. This leads to $g_1(T_\text{max})$ being noticeably smaller than $g_\text{max}$. Furthermore, the exact results show a small revival of the maximum at $t\approx 160 J^{-1}$. In contrast, the TD2RDM results lack this shift, resulting in an overestimation of the amplitude $g_1(T_\text{max})$. Additionally, the TD2RDM results exhibit prominent periodic revivals, indicating that relaxation effects are underestimated. Notably, the TD2RDM prediction for $U = 4J$ is remarkably accurate, even capturing the third revival of $g_1(t)$.\\
This analysis can be complemented by means of the entropy $S_2(t)$, which concentrates the information on all geminal occupation numbers into one quantity. We observe that $S_2(t)$ increases as a function of time, reaching $S_2^*$ for an extreme AGP state at a time close to $T_\text{max}$ [Fig.~\ref{fig:N8_entropies_cumulants} (a)]. The TD2RDM results closely follow the exact results initially, but then $S_2(t)$ remains large within the exact results while spuriously fluctuating within the TD2RDM results. Another quantity that gives insights into pair-correlations is the two-particle cumulant Eq.~\ref{eq:cumu2}. Evaluating the Frobenius norms of the two-particle cumulants for spin-polarized pairs, $|\Delta_{12}^{\uparrow\uparrow}|_2$, and spin-unpolarized pairs $|\Delta_{12}^{\uparrow\downarrow}|_2$ as a measure for two-particle correlations reveals that the TD2RDM overestimates the pairing of spin-unpolarized pairs while underestimating the production of spin-polarized pairs, see Fig.~\ref{fig:N8_entropies_cumulants} (b) and (c).\\
We, therefore, focus in our further analysis on the scaling behavior of $T_\text{max}$ and $g_1(T_\text{max})$, which is overall very well captured by the TD2RDM method, see Figs.~\ref{fig:N8_Tmax} and \ref{fig:N8_amp_Tmax}.
We would like to point out that our previous results \cite{donsa_nonequilibrium_2023} indicate that the TD2RDM method is more accurate for larger systems with $M_s\gtrsim 20$ than for systems as small as $M_s = 8$. While for $M_s=8$ the TD2RDM method overestimates the density fluctuations for $t> 60 J^{-1}$ (not shown) consistent with the deviations obtained for the geminal occupation numbers, we have observed almost perfect agreement for the density fluctuations in case of $M_s = 18$ in \cite{donsa_nonequilibrium_2023}. The predictions for $T_\text{max}$ and the amplitude $g_1(T_\text{max})$ within TD2RDM might, therefore, be even more accurate than a straight forward extrapolation of the results for $M_s=8$ allows to judge.
\subsection{Larger systems in 1D}\label{subsec:larger}
\begin{figure}[t]
	\includegraphics[width=\columnwidth]{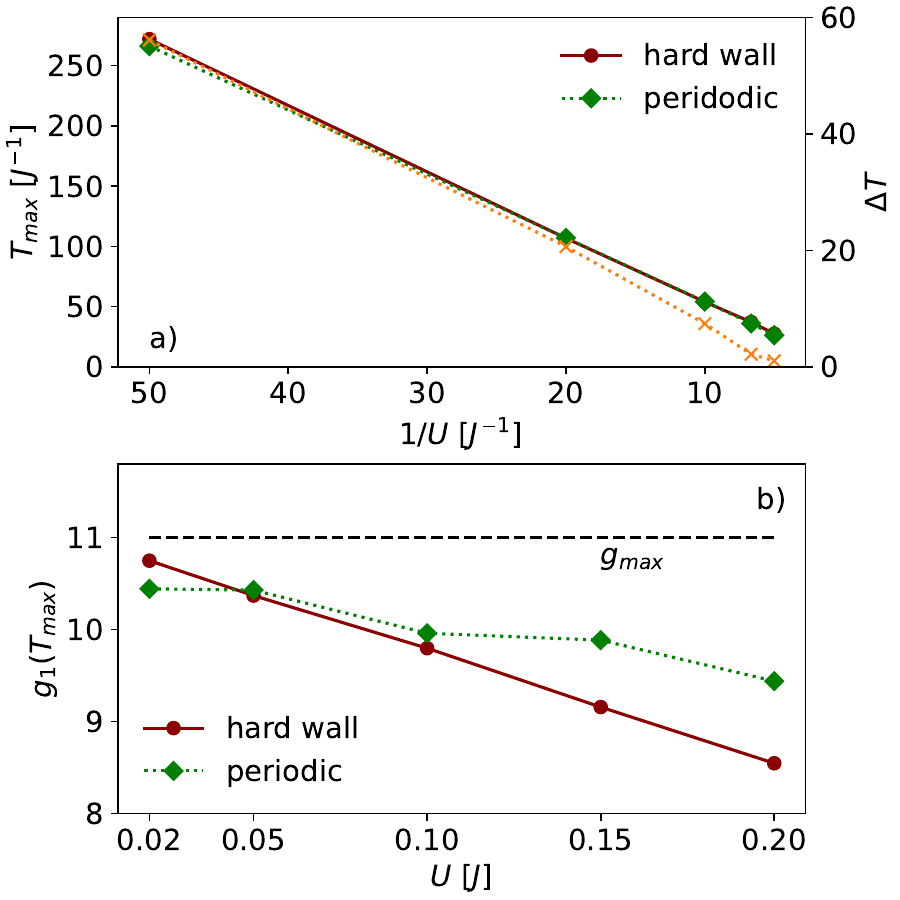}
	\caption{(a) $T_\text{max}$ and (b) the amplitude of the maximum of the largest geminal occupation number $g_1(T_\text{max})$ as a function of $U$ for $M_s=20$ sites and different boundary conditions. In (a) on the second y-axis we plot with crosses exemplary for hard wall boundary conditions $\Delta T$ (i.e.~the width of the time window, where the density fluctuations become homogeneous).}
	\label{fig:N20_amp_Tmax}
\end{figure}

We now turn to the fundamental question of whether the observed effect exhibits characteristics of a quasicondensate that vanishes in the thermodynamic limit, or whether this dynamical setting enables a circumvention of the Mermin-Wagner-Hohenberg theorem which prohibits condensation in 1D in equilibrium \cite{hohenberg_existence_1967}. To address this question, we extend our investigation using the TD2RDM method to explore significantly larger system sizes for which there are no exact benchmarks available, bearing in mind that the TD2RDM might slightly overestimate both $T_\text{max}$ as well as $g_\text{max}$ while maintaining a reliable prediction of the overall behavior and scaling.\\
We start our investigation with a system of $M_s = 20$ for which we make detailed comparison to the system with $M_s = 8$.
\begin{figure}[t]
	\includegraphics[width=\columnwidth]{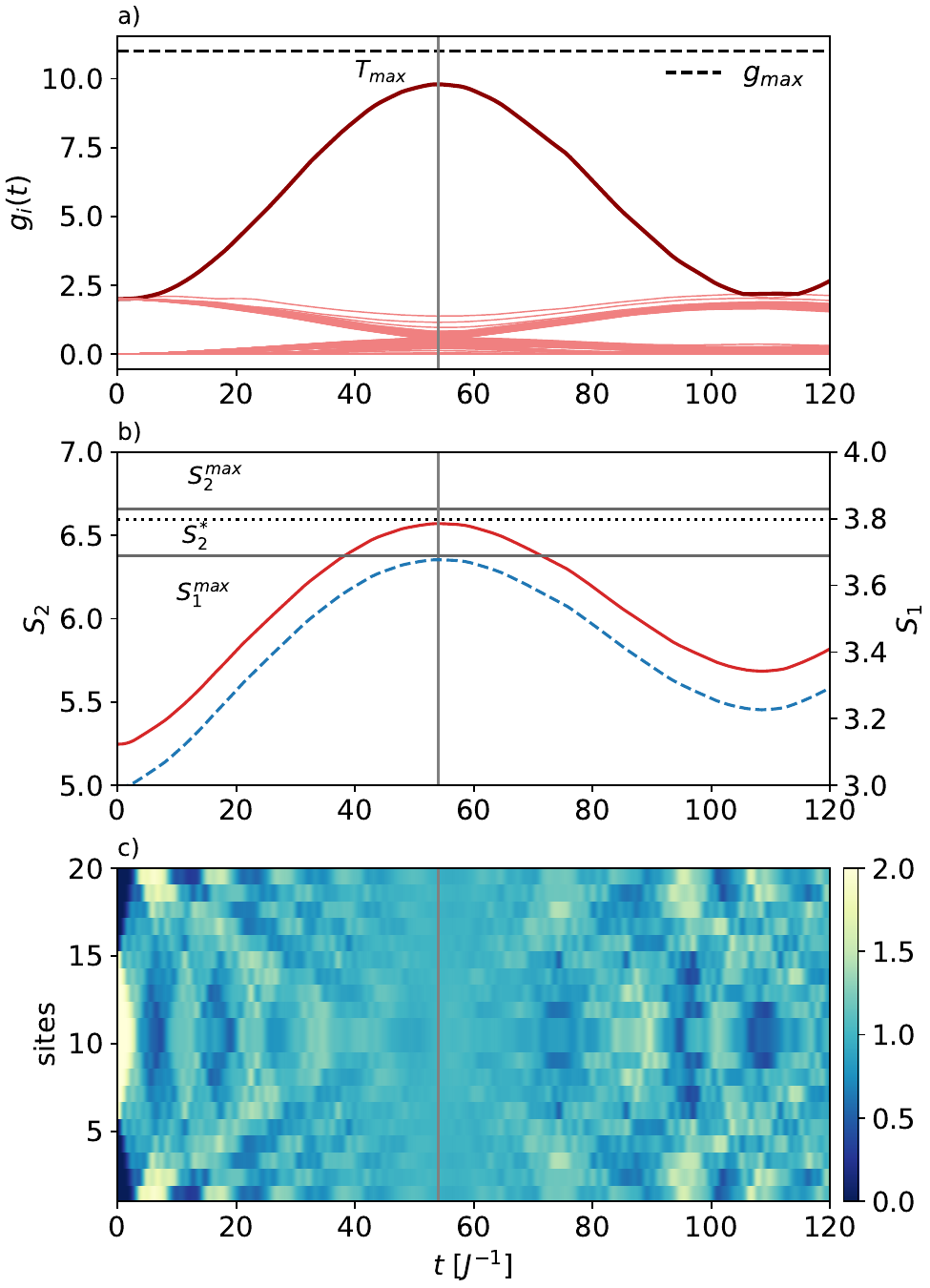}
	\caption{TD2RDM results for the Fermi-Hubbard model in 1D with $M_s = 20$, $U=0.1 J$, hard-wall boundary conditions, and a centered initial condition. (a) Geminal occupation numbers, (b) entropies $S_2$ (left y-axis) and $S_1$ (dashed line, right y-axis), and (c) particle-density fluctuations as a function of time. The vertical lines mark the time $T_\text{max}$. In (b) the upper solid horizontal line corresponds to $S_2^\text{max}$, the lower corresponds to $S_1^\text{max}$, and the dotted horizontal line corresponds to $S_2^*$.}
	\label{fig:N20_all_gems_den}
\end{figure}
Fig.~\ref{fig:N20_all_gems_den} shows one particular example of the behavior for $M_s= 20$. Similar to the case of $M_s=8$, the largest geminal occupation number $g_1(t)$ rises to a pronounced maximum of $g_1(T_\text{max})\approx 10$ close to $g_\text{max} = 11$ and several revivals appear [Fig.~\ref{fig:N20_all_gems_den} (a)]. These strong revivals are again most likely spurious.\\
It is instructive to explore in parallel the density fluctuations of the system [Fig.~\ref{fig:N20_all_gems_den} (c)]. We observe strong fluctuations in between the boundaries of the system and several reflections from the hard walls except for a short interval in time around $T_\text{max}$, where the particle density becomes homogeneous. This effect is accompanied by a local maximum in the single-particle entropy $S_1$ obtained from the diagonalization of the 1RDM (Eq.~\ref{eq:1rdm})
\begin{equation}
	D_1(t) = \sum_{j=1}^{r} n_j(t)|n_j(t)\rangle \langle n_j(t)|,
	\label{eq:1rdm_diag}
\end{equation}
given by
\begin{equation}
	S_1(t) = -\sum_{j=1}^{r} n_j(t)\ln{n_j(t)} .
	\label{eq:S1}
\end{equation}
We have again renormalized $\sum_{i=1}^rn_i = 1$ to calculate the entropy $S_1$.
The maximum at $T_\text{max}$ is given by $S_1(T_\text{max}) = 3.68$, which is very close to the maximal value of $S_1^\text{max} = 3.69$ for equally distributed natural occupation numbers for $r = 2M_s=40$ [see Fig.~\ref{fig:N20_all_gems_den} (b)]. Overall, $S_1(t)$ follows the curve of the two-particle entropy $S_2(t)$.\\
The homogeneous distribution of the particle density over the entire system follows from the fact that an extreme AGP state belongs to one of the quantum many-body states that maximize $S_1$, and is a direct consequence of the fact that the observed state is close to an extreme AGP state, with natural orbitals distributed over the entire system and their occupation numbers being almost equal. This striking effect of the particle density becoming homogeneous during a small but finite time interval $\Delta T$ around $T_\text{max}$ opens the door for an experimental study of these effects within the platforms of ultracold quantum simulators. While the increase of the pair-state occupation number is not easily accessible experimentally, the monitoring of the density fluctuations of the system has become an experimental routine on these platforms (see e.g.~\cite{cheuk_quantum-gas_2015, parsons_site-resolved_2015, haller_single-atom_2015, edge_imaging_2015, omran_microscopic_2015}, for a current experimental realization of a 1D periodic system see \cite{cai_persistent_2022}) and would allow to measure $T_\text{max}$ for even larger systems and other geometries. $\Delta T$ depends on the width of the maximum in $g_1(t)$, which increases with $T_\text{max}$, i.e.~with $1/U$ [see Fig.~\ref{fig:N20_amp_Tmax} (a)]. In Fig.~\ref{fig:N20_amp_Tmax} (a) we plot $\Delta T$ exemplary for hard wall boundary conditions by determining the time interval for which the weight of the Fourier components with $k>0$ for the density fluctuations in Fig.~\ref{fig:N20_all_gems_den} (c) fall below a certain limit. \\
When considering different boundary conditions and positions of the initial state, we observe again only a weak dependence of the dynamical (quasi-)condensation effect on them. $T_\text{max}$ as a function of $U$ is practically equal for both boundary conditions, see Fig.~\ref{fig:N20_amp_Tmax} (a).
\begin{figure*}[t]
	\includegraphics[width=14cm]{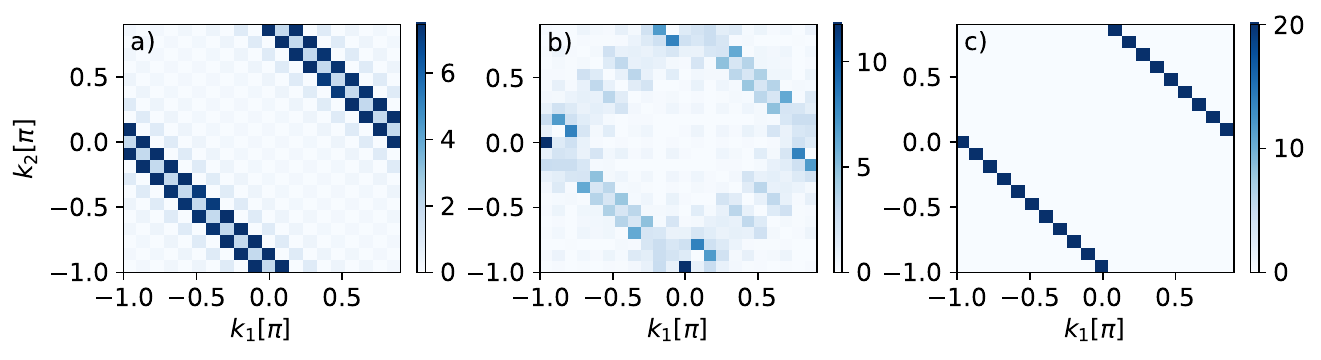}
	\caption{Absolute square of the momentum distribution of the state $|g_1(T_\text{max})\rangle$ for $M_s = 20$, $U=0.1 J$ and (a) periodic boundary conditions, and (b) hard-wall boundary. (c) Absolute square of the momentum distribution of the $\eta$-condensate for $M_s = 20$ and periodic boundary conditions.}
	\label{fig:N20_state_analysis}
\end{figure*}
$g_1(T_\text{max})$ behaves similarly for the two different boundary conditions but overall the decay with $U$ is slower in case of periodic boundary conditions and less monotonic. 
A further analysis of the state $|g_1(T_\text{max})\rangle$ reveals another interesting consequence of the boundary conditions. When comparing the state $|g_1(T_\text{max})\rangle$ to the $\eta$-condensate in momentum space we observe for periodic boundary conditions that $|g_1(T_\text{max})\rangle$ shows the typical $\eta$-pairing of $k$ and $k-\pi$ for particles of different spin, see Fig.~\ref{fig:N20_state_analysis}. $|g_1(T_\text{max})\rangle$ shows only small deviations from the $\eta$-condensate [see Fig.~\ref{fig:N20_state_analysis} (c)] most notably a larger spread and a local minimum across the $k_2 = \pi - k_1$ lines in momentum space [Fig.~\ref{fig:N20_state_analysis} (a)]. In contrast, for the system with hard-wall boundary conditions, the reflexions at the boundaries lead to the emergence of additional pairings, most notably Cooper-pair like pairs with $k_1 = -k_2$, see Fig.~\ref{fig:N20_state_analysis} (b). Note that the absolute square of the momentum distribution for the $\eta$-state is practically equal for periodic and hard-wall boundary conditions, the only difference being a different phase of the amplitudes and two more data points in Fourier space due to explicitly taking into account the vanishing density at site number $0$ and $M_s +1$ for hard-wall boundary conditions.
Despite these differences in the properties of the (quasi-)condensate state, the geminal occupation numbers show similar behavior for both boundary conditions [Fig.~\ref{fig:N20_amp_Tmax}].\\
We now turn to analyzing the behavior of the (quasi)-condensation effect with system size.
\begin{figure}[t]
	\includegraphics[width=\columnwidth]{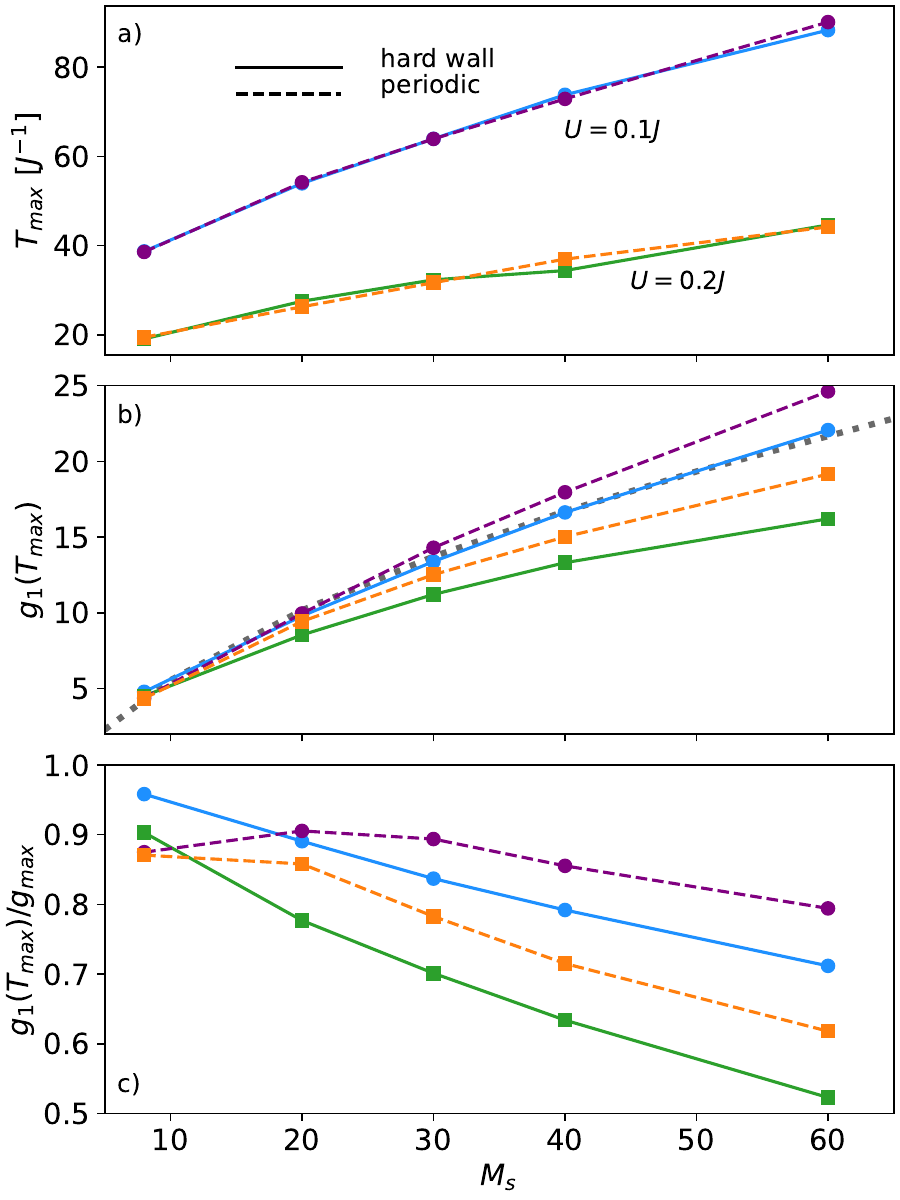}
	\caption{Dynamical quasi-condensation as functions of system size $M_s$. (a) $T_\text{max}$, (b) $g_1(T_\text{max})$, and (c) $g_1(T_\text{max})$ relative to the maximum $g_\text{max}$ (Eq.~\ref{eq:gmax_gen}) for hard-wall boundary conditions (solid), periodic boundary conditions (dashed), $U=0.1J$ (dots), and $U=0.2J$ (squares). The gray dotted line in (b) corresponds to a fit to a square root function.}
	\label{fig:N_large_amp_Tmax}
\end{figure}
Since parameter scans are numerically increasingly demanding with system size we focus on two parameters for the on-site interaction, i.e.~$U = 0.1 J$ and $U = 0.2 J$.
%
%
We observe for both values of $U$ that $T_\text{max}$ increases with $M_s$ in a non-linear way [Fig.~\ref{fig:N_large_amp_Tmax} (a)]. The influence of the boundary conditions is negligible. In agreement with the previous investigations for $M_s = 8$ we observe that $T_\text{max}$ is larger for $U=0.1J$ than for $U=0.2J$ for all investigated $M_s$. The amplitude of the first maximum $g_1(T_\text{max})$ increases monotonically with $M_s$ [Fig.~\ref{fig:N_large_amp_Tmax} (b)]. Most importantly, in the given available interval, the increase is proportional to $\sqrt{M_s}$. Consequently, we observe a monotonic decrease of $g_1(T_\text{max})$ with respect to $g_\text{max}$. Since $g_\text{max}$ grows linearly with $M_s$ (Eq.~\ref{eq:gmax_gen}) our results indicate that $g_1(T_\text{max})/g_\text{max} \propto 1/\sqrt{M_s}$ in the thermodynamic limit. In other words, if the $\sqrt{M_s}$-dependence of $g_1(T_\text{max})$ is preserved in the thermodynamic limit of $M_s\rightarrow \infty$, then in the thermodynamic limit, $g_1(T_\text{max})/g_\text{max}\rightarrow 0$. Our results thus predict that the observed effect is a dynamical quasi-condensate.\\
In the following section, we will investigate whether we might obtain a true condensation in 2D, i.e.~in the presence of transverse degrees of motion, by extending the system to narrow 2D stripes. 
\subsection{Extension to 2D systems}\label{subsec:2d}
\begin{figure}[t]
	\includegraphics[width=\columnwidth]{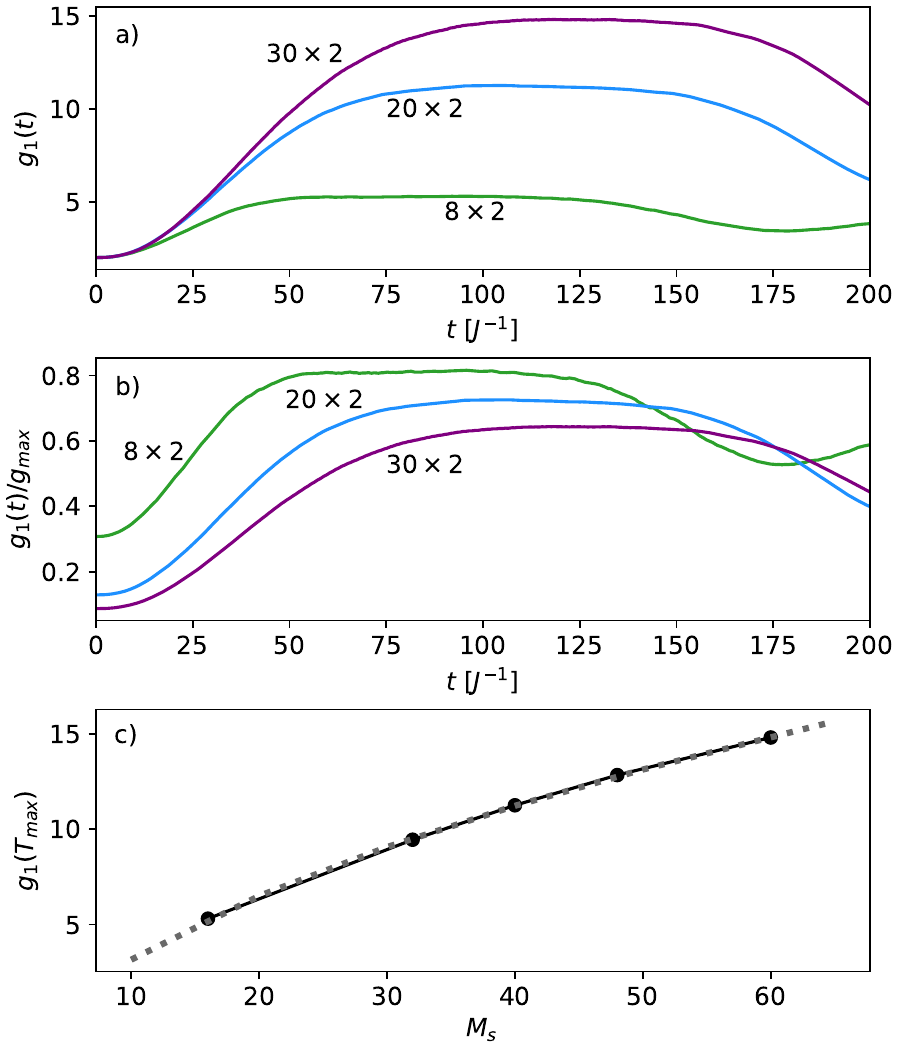}
	\caption{(a) The largest geminal occupation number $g_1(t)$, and (b) $g_1(t)$ relative to $g_\text{max}$ as a function of time for $U=0.1J$ for different 2D stripe geometries with $M_s = 8\times 2$ and $N=8$, $M_s = 20\times 2$ and $N=20$, and $M_s = 30\times 2$ and $N=30$. (c) The largest value $g_1(T_\text{max})$ as a function of the number of sites $M_s$ for the systems in (a). The gray dashed line represents a fit to square root function of $M_s$.}
	\label{fig:2D_2}
\end{figure}

\begin{figure}[t]
	\includegraphics[width=\columnwidth]{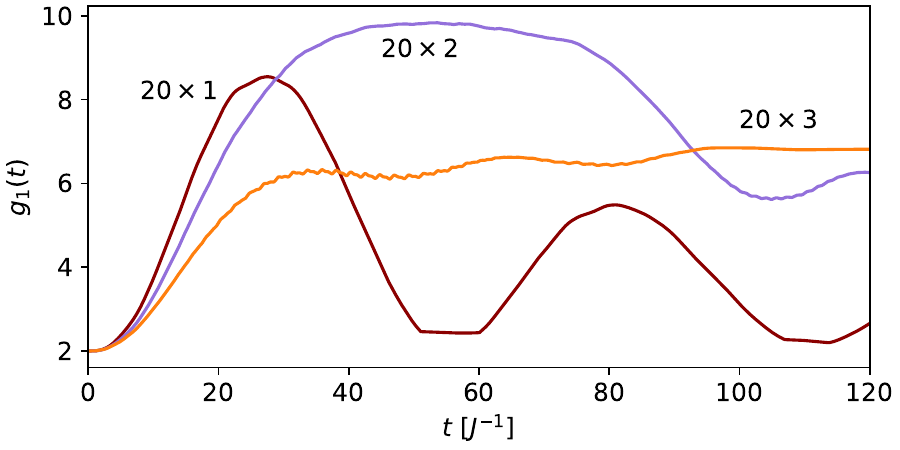}
	\caption{The largest geminal occupation number $g_1(t)$ as a function of time for $U=0.2J$ and number of particles $N=20$ and different stripe geometries with $M_s = 20\times 1$ (1D), $M_s = 20\times 2$, and $M_s = 20\times 3$ sites amounting to different fillings.}
	\label{fig:2D_1_3}
\end{figure}

We restrict our study here to hard-wall boundary conditions as these are more easily realizable experimentally in 2D systems. The geometries we investigate here are depicted in Fig.~\ref{fig:overview} (b) and (c). For these systems with large $M_s$ systematic scans with $U$ and $M_s$ are increasingly expensive even within TD2RDM such that we focus here on the question whether we get an indication of a convergence to finite values of $g_1(T_\text{max})/g_\text{max}$ in the thermodynamic limit. For this purpose we stick to $U=0.1 J$ and $U=0.2 J$ for different geometries and numbers of sites $M_s$. \\
When extending the 1D system by one transversal site along the entire length as in Fig.~\ref{fig:overview} (b) we observe a build-up of a maximum in $g_1(t)$ over time, see Fig.~\ref{fig:2D_2} (a) for different $M_s$ and a filling of $1/4$ (i.e.~$N=M_s/2=r/4$), and Fig.~\ref{fig:2D_2} (b) for $g_1(t)$ relative to $g_\text{max}$. 
In this case, the prediction for $g_\text{max}$ (Eq.~\ref{eq:gmax_gen}) is equal to $g_\text{max} = (3N+2)/4$, i.e.~larger than in the case of half-filling, Eq.~\ref{eq:gmax_halff}. It is striking that the maxima are much broader than in 1D extending over longer periods of time. For $g_1(T_\text{max})$ we observe a sub-linear (square root) increase with $M_s$ [Fig.~\ref{fig:2D_2} (c)], while convergence of $g_1(T_\text{max})/g_\text{max}$ towards finite values would require a linear increase. The conclusion from these results is thus that the condensation effect will vanish in the thermodynamic limit.\\
Increasing the system further by another transverse degree of freedom as in Fig.~\ref{fig:overview} changes the picture quite strongly, see Fig.~\ref{fig:2D_1_3}. For the system with $M_s = 20 \times 3$ sites we observe a fast increase in $g_1(t)$ initially but then a flattening and a further increase at a much smaller pace. Until a time of around $t=100 J^{-1}$, which is already numerically quite costly, the largest geminal occupation number $g_1$ does not come close to $g_\text{max} = (5N+2)/6 =17 $ for this case of sixth filling. It is difficult to estimate based on these observations whether the system develops a (quasi-)condensate at some later point in time but the curves obtained so far indicate that the effect does not persist in 2D.
%
\section{Conclusions}\label{sec:concld}
We have numerically investigated dynamical quasi-condensation in the Fermi-Hubbard model starting from a completely uncorrelated initial state of adjacent doubly occupied sites. In 1D we have shown that upon expansion of the system the largest eigenvalue of the two-particle reduced density matrix (2RDM), $g_1(t)$, develops a local maximum that comes close to the theoretical upper limit $g_\text{max}$ predicted by Yang \cite{yang_concept_1962}, signaling the appearance of fermionic pair condensation. This dynamical quasi-condensation is accompanied by strong two-particle correlations as measured by the two-particle cumulants and the two-particle entropy obtained from the eigenstates of the 2RDM. This condensation effect appears for all investigated values of the interaction $U$, but shows a distinctly different behavior for small $U<1 J$ as compared to large $U>1J$, where $J$ is the hopping matrix element. In the case of $U> 1J$, the quasi-condensation effect has been explained by the physics of an emergent Hamiltonian \cite{vidmar_emergent_2017}, and requires a completely free expansion. In contrast, for small $U<1J$ we observe that the system undergoes many interactions with the boundaries of the system on the characteristic time scale $T_\text{max}$ during which the dynamical quasi-condensate emerges. The effect is only weakly dependent on whether we chose hard-wall or periodic boundary conditions. Interestingly, the two-particle quasi-condensate state does dependent on the boundary conditions and features the typical $\eta$-condensate pairing in case of periodic boundary conditions and both an $\eta$-condensate pairing and Cooper-pair-like pairing in case of hard-wall boundary conditions. Moreover, the scaling of both $T_\text{max}$ and the amplitude $g_1(T_\text{max})$ with $U$ is markedly different for $U<1 J$ and $U>1J$. For $U<1J$ we observe that $T_\text{max}\propto 1/U$, while for $U>1J$ one obtains $T_\text{max}\propto U$.\\
It is so far an open question whether it is possible to circumvent the Mermin-Wagner-Hohenberg theorem \cite{mermin_absence_1966, hohenberg_existence_1967} in a dynamical setting, which prohibits the appearance of condensation in 1D in the thermodynamic limit in equilibrium. We address this question for our weakly interacting quasi-condensation effect by expanding to system sizes of up to $60$ sites. To propagate these systems over the required long periods of time we employ our newly developed time-dependent 2RDM (TD2RDM) method. Based on comparisons with exact results for small system sizes we have demonstrated that the TD2RDM method accurately predicts the essential features of the effect such that extrapolation to large systems, where exact benchmarks do not exist, can be made. The analysis of the effect with increasing system size reveals that the maximum of $g_1$ grows proportionally to the square root of the system size while $g_\text{max}$ grows linearly. Our results thus indicate that $g_1(T_\text{max})/g_\text{max}$ vanishes in the thermodynamic limit, a defining feature of a quasi-condensate.\\ 
To further expand on this question, we have extended our system to narrow 2D stripes allowing for transversal motion during expansion. For systems with two transverse sites, we observe broad maxima in $g_1$ over longer periods of time compared to the 1D case. The scaling of $g_1(T_\text{max})$ with the system size, however, again indicates a vanishing condensation effect in the thermodynamic limit. A further increase, of the transverse degrees of freedom seems to lead to a strong deterioration of the effect. In this case, $g_1$ increases strongly initially but then flattens, featuring a slow but steady increase over the entire investigated time interval during which, however, $g_1$ does not come as close to $g_\text{max}$ as in the previous cases. Further increase in the number of transversal degrees of freedom to approach a sufficiently large square lattice is planned in future.\\ 
Our results open the door to further scrutinize this effect in the platform of experimental quantum simulators with single-site resolution \cite{cheuk_quantum-gas_2015, parsons_site-resolved_2015, haller_single-atom_2015, edge_imaging_2015, omran_microscopic_2015} where substantially larger system sizes both in 1D and 2D could be probed. We have shown that the appearance of the quasi-condensation effect is accompanied with spatial particle density fluctuations becoming homogeneous over the entire system for a short but finite period of time around $T_\text{max}$. Strong density fluctuations reappear shortly after $T_\text{max}$. This effect can be traced back to the single-particle entropy obtained from the eigenvalues of the one-particle reduced density matrix showing a local maximum. Since probing the particle density distribution is an experimental routine nowadays, observing a homogeneous density over finite periods of time could serve as a strong indication of the dynamical (quasi-)condensation effect. \\
Finally, our results highlight the potential of the TD2RDM method to deliver predictions for system sizes and time scales not reachable by any other method. We would like to point out that our present code, while exploiting several symmetries of the system, is not fully optimized and heavily parallelized yet. Further improvement of its optimization and employing multi-node parallelization should allow to compute the dynamics in systems with a number of sites of about a factor of two larger than the largest systems in the present study over similar time scales. With this extension, investigations of condensation effects on a square lattice in 2D should become possible. The TD2RDM method could thus provide benchmarks and guidance for studies on non-equilibrium systems on quantum simulators for large systems.
\section*{Acknowledgements}
We thank Slavatore R. Manmana and Joseph Tindall for helpful discussions and hints on literature. IB thanks the Simons Foundation for the great hospitality and support during her research visit at the CCQ of the Flatiron Institute, where parts of this research were conducted. The Flatiron Institute is a division of the Simons Foundation. We acknowledge support from the Max Planck-New York City Center for Non-Equilibrium Quantum Phenomena, 
Cluster of Excellence `CUI: Advanced Imaging of Matter'- EXC 2056 - project ID. This research was funded by the Austrian Science Fund (FWF) grant P 35539-N. Calculations were performed on the Vienna Scientific Cluster (VSC4). 
\appendix
\section{Conservation of $\eta$-symmetry within purification and details on the numerical implementation}\label{app:eta_sym}
In order to  preserve N-representability of the 2RDM at least partially during time propagation, we apply an a-posteriori purification procedure after propagation time steps \cite{lackner_propagating_2015, lackner_high-harmonic_2017, joost_dynamically_2022, donsa_nonequilibrium_2023}. Without purification the TD2RDM method tends to produce 2RDMs with negative eigenvalues that ultimately may lead to instabilities \cite{akbari_challenges_2012}. Purification entails removing iteratively the defective part from the 2RDM while preserving its contraction to the 1RDM. This can be facilitated by applying the unitary decomposition of the 2RDM \cite{mazziotti_purification_2002},
\begin{equation}
	D_{12} = D_{12;\perp}\{D_1\} + D_{12;K},
	\label{eq:d12_ud}
\end{equation}
where the kernel $D_{12; K}$ has vanishing traces, and all the trace information is contained in the orthogonal component, $D_{12,\perp}$. $D_{12,\perp}\{D_1\}$ indicates that the orthogonal component is a functional of the 1RDM. We have empirically shown \cite{lackner_propagating_2015, lackner_high-harmonic_2017, donsa_nonequilibrium_2023} that restoring the positive-semidefinitness of the 2RDM and the two-hole RDM is sufficient to stabilize the equations of motion in most cases \cite{donsa_nonequilibrium_2023}. Accordingly, we determine the defective part of the 2RDM
\begin{equation}
	D_{12}^< = \sum_{g_i<0} g_i |g_i\rangle \langle g_i|
\end{equation}
and similarly the defective part of the two-hole RDM, $Q_{12}^<$. Subtracting the kernel of the defective components of both the 2RDM and the two-hole RDM from the 2RDM iteratively leads to a positive semi-definite 2RDM, while preserving contraction consistency to the 1RDM. In general, however, other symmetries like e.g.~energy conservation are broken in this process unless further amendments are applied. It has been shown \cite{joost_dynamically_2022} that energy conservation can be restored by removing from $D_{12}^<$ and $Q_{12}^<$ the contributions that would lead to violations of energy conservation, before calculating the kernel. Similarly, to enforce conservation of $\langle \hat \eta^+\hat\eta-\rangle$, Eq.~\ref{eq:eta_consv}, we set ${D^<}_{j\uparrow j \downarrow}^{i\uparrow i\downarrow}$ and ${Q^<}_{j\uparrow j \downarrow}^{i\uparrow i\downarrow}$to zero for all $i$ and $j$. We then determine the kernel of the correspondingly corrected defective components $D_{12}^<$ and $Q_{12}^<$, and subtract them from the 2RDM. This procedure is iterative repeated until the required threshold for the smallest geminal occupation number is reached, or until the selected maximal number of iterations is reached.\\
For the studies in the main text we have applied a threshold of $0$ for the smallest geminal occupation number and at most $10$ iteration steps within the purification for all systems except for $M_s=60$, where only one step is applied due to its scaling with $M_s^6$ within the current implementation using full diagonalization of the 2RDM. Purification is applied after each global time step of $dt = 0.02 J^{-1}$ within which the equations of motion, Eq.~\ref{eq:eom}, are solved with adaptive time steps using a Runge-Kutta-Fehlberg propagator of $4^{th}$ and $5^{th}$ order. The different orders are used to adapt the time steps to achieve a prescribed accuracy. We have checked that changing the global time step to $dt = 0.01 J^{-1}$, i.e.~applying up to twice as many purification steps, leads to the same results to high degree of accuracy.
\section{Formalism of emergent Hamiltonians}\label{app:emer_hamil}
We explore the applicability of the formalism of emergent Hamiltonians to the dynamical quasi-condensation effect for $U<1$. Following \cite{vidmar_emergent_2017}, we define a Hamiltonian $\hat H_0$, such that the initial state is an eigenstate of $\hat H_0$. One particular choice following Ref.~\cite{vidmar_emergent_2017} for the present initial condition is
\begin{equation}
    \hat H_0 = \frac{1}{M_s}\sum_{j=1}^{M_s} j(\hat n_j^\uparrow + \hat n_j^\downarrow).
\end{equation}
There are, however, other possible choices such as $\hat H_0 = U\sum_{j=1}^{M_s} \hat n_j^\uparrow \hat n_j^\downarrow$, but they lead to essentially the same conclusions, see below. The system evolves under the influence of $\hat H$, Eq.~\ref{eq:ham}, such that $\hat P = \hat H_0-\hat H$ can be regarded as a quench. Starting from the initial condition $(\hat H_0 -\lambda)|\Psi(0)\rangle = 0$ one applies the time evolution operator to obtain 
\begin{equation}
    \left(
    e^{-i\hat H t}\hat H_0 e^{i\hat H t}-\lambda \right)|\Psi(t)\rangle := \hat M(t)|\Psi(t)\rangle.
\end{equation}
The operator $\hat M(t)$ is given by
\begin{equation}\label{eq:M}
\begin{split} 
    \hat M(t) & = \hat H - \lambda + \hat P  -it [\hat H,\hat P] + \frac{(-it)^2}{2} \left[ \hat H,[\hat H, \hat P]\right] + ...  \\
    & = \hat H -\lambda +\hat P  -it\hat Q
    + \sum_{n=1}^{\infty} \frac{(-it)^{n+1}}{(n+1)!} 
    \mathcal{\hat H}_n,
\end{split}
\end{equation}
with the definition $\hat Q = [\hat H, P]$, and $\mathcal{\hat H}_n$ describing nested commutators starting with $\mathcal{\hat H}_1 = [\hat H,\hat Q]$.
According to \cite{vidmar_emergent_2017} the emergent Hamiltonian is applicable if $\hat M(t)$ is a local operator. One particular family studied corresponds to $[\hat H,\hat Q] = 0$ up to boundary terms, where the time propagated state is exponentially close to the eigenstate of the emergent Hamiltonian for times proportional to the system size \cite{vidmar_emergent_2017}.\\
For the following discussion it is convenient to separate our Hamiltonian Eq.~\ref{eq:ham} as $\hat H = \hat T +\hat W$, where $\hat T$ stands for the hopping operator proportional to $J$ and $\hat W$ is the on-site interaction proportional to $U$. For our system with hard wall boundary conditions (similar results hold for periodic boundary conditions) we obtain
\begin{equation}
\begin{split}
    \hat Q = [\hat H, \hat P ] = [\hat T, \hat P ] &= \frac{J}{M_s}\sum_{j=1}^{M_s-1}\sum_\sigma \hat a_{j+1\sigma}^\dagger \hat a_{j\sigma} \\
    &-\frac{J}{M_s}\sum_{j=1}^{M_s-1}\sum_\sigma \hat a_{j\sigma}^\dagger \hat a_{j+1\sigma}
\end{split}
\end{equation}
(note that with the prefactor $(-i)$ from Eq.~\ref{eq:M} this contribution to $\hat M(t)$ is Hermitian as it should be), and 
\begin{equation}
\begin{split}       
    \mathcal{\hat H}_1 = [\hat H,\hat Q]  = \frac{2J^2}{M_s}&\left(
    n_{M_s}^\uparrow + n_{M_s}^\downarrow
    - n_{1}^\uparrow - n_{1}^\downarrow
    \right)  \\
     + \frac{UJ}{M_s} &\Bigg[
    \sum_{j=1}^{M_s-1} \hat a_{j+1\uparrow }^\dagger a_{j\uparrow }\left(\hat n_{j+1}^{\downarrow} - \hat n_{j}^{\downarrow}
    \right) + \text{H.c.} \\
    & + \sum_{j=1}^{M_s-1} \hat a_{j+1\downarrow}^\dagger a_{j\downarrow}\left(\hat n_{j+1}^{\uparrow} - \hat n_{j}^{\uparrow}
    \right) + \text{H.c.}\Bigg].
\end{split}
\end{equation}
The first contribution comes from the commutator between $\hat T$ and $\hat Q$ (and corresponds to the one found in \cite{vidmar_emergent_2017} for the non-interacting system), while the second term comes from the commutator between $\hat W$ and $\hat Q$. Clearly, the commutator $[\hat  H, \hat Q]$ does not vanish and leads to a term growing quadratically in time in $\hat M(t)$, Eq.~\ref{eq:M}. Higher orders $\mathcal{\hat H}_n$ contain elements of the form 
\begin{equation}\label{eq:M_higher_order}
\begin{split}       
    \frac{U^nJ}{M_s}&\Bigg[
    \sum_{j=1}^{M_s-1} \hat a_{j+1\uparrow }^\dagger a_{j\uparrow }\left(\hat n_{j+1}^{\downarrow} - \hat n_{j}^{\downarrow}
    \right)^n + (-1)^{n+1}\text{H.c.} \\
    & + \sum_{j=1}^{M_s-1} \hat a_{j+1\downarrow}^\dagger a_{j\downarrow}\left(\hat n_{j+1}^{\uparrow} - \hat n_{j}^{\uparrow}
    \right)^n + (-1)^{n+1}\text{H.c.}\Bigg],
\end{split}
\end{equation}
which originate from the evaluation of the $n$ nested commutators $[\hat W, [\hat W,\hdots,[\hat W,\hat Q]]]$. In addition, $\mathcal{\hat H}_n$
contains all lower orders in $U$, i.e.~$U^j$ with $j\in[0,n-1]$ originating from different sequences of $\hat T$ and $\hat W$ in the nested commutators and leading to products of $2(j+1)$ creation and annihilation operators. These products contain elements of the form $a_j^\dagger a_{j+m}$ with $m\leq n$ (up to $m=M_s-1$, i.e.~up to the span over the entire system). The contribution independent of $U$ originates from repeated application of the commutator with $\hat T$ and contains only boundary terms. $\mathcal{M}(t)$ therefore contains arbitrary high orders of products of creation and annihilation operators with an increasing span over the entire system, albeit being suppressed by $U^m$ with $U<1$ in our case and $m$ the number of commutators containing $\hat W$. While the contributions from Eq.~\ref{eq:M_higher_order} to Eq.~\ref{eq:M} can be evaluated to all orders leading to an extensive sum of local exponential operators, the other terms amount to an infinite series of non-local operators and thus cannot be easily treated. These results underline our conclusion that the present effect is not captured by an emergent Hamiltonian.
\newpage
\section*{References}

\begin{thebibliography}{53}%
	\makeatletter
	\providecommand \@ifxundefined [1]{%
		\@ifx{#1\undefined}
	}%
	\providecommand \@ifnum [1]{%
		\ifnum #1\expandafter \@firstoftwo
		\else \expandafter \@secondoftwo
		\fi
	}%
	\providecommand \@ifx [1]{%
		\ifx #1\expandafter \@firstoftwo
		\else \expandafter \@secondoftwo
		\fi
	}%
	\providecommand \natexlab [1]{#1}%
	\providecommand \enquote  [1]{``#1''}%
	\providecommand \bibnamefont  [1]{#1}%
	\providecommand \bibfnamefont [1]{#1}%
	\providecommand \citenamefont [1]{#1}%
	\providecommand \href@noop [0]{\@secondoftwo}%
	\providecommand \href [0]{\begingroup \@sanitize@url \@href}%
	\providecommand \@href[1]{\@@startlink{#1}\@@href}%
	\providecommand \@@href[1]{\endgroup#1\@@endlink}%
	\providecommand \@sanitize@url [0]{\catcode `\\12\catcode `\$12\catcode
		`\&12\catcode `\#12\catcode `\^12\catcode `\_12\catcode `\%12\relax}%
	\providecommand \@@startlink[1]{}%
	\providecommand \@@endlink[0]{}%
	\providecommand \url  [0]{\begingroup\@sanitize@url \@url }%
	\providecommand \@url [1]{\endgroup\@href {#1}{\urlprefix }}%
	\providecommand \urlprefix  [0]{URL }%
	\providecommand \Eprint [0]{\href }%
	\providecommand \doibase [0]{https://doi.org/}%
	\providecommand \selectlanguage [0]{\@gobble}%
	\providecommand \bibinfo  [0]{\@secondoftwo}%
	\providecommand \bibfield  [0]{\@secondoftwo}%
	\providecommand \translation [1]{[#1]}%
	\providecommand \BibitemOpen [0]{}%
	\providecommand \bibitemStop [0]{}%
	\providecommand \bibitemNoStop [0]{.\EOS\space}%
	\providecommand \EOS [0]{\spacefactor3000\relax}%
	\providecommand \BibitemShut  [1]{\csname bibitem#1\endcsname}%
	\let\auto@bib@innerbib\@empty
	\bibitem [{\citenamefont {Abrikosov}\ \emph {et~al.}(1965)\citenamefont
		{Abrikosov}, \citenamefont {Gor'kov},\ and\ \citenamefont
		{Dzyaloshinskii}}]{abrikosov_quantum_1965}%
	\BibitemOpen
	\bibfield  {author} {\bibinfo {author} {\bibfnamefont {A.~A.}\ \bibnamefont
			{Abrikosov}}, \bibinfo {author} {\bibfnamefont {L.~P.}\ \bibnamefont
			{Gor'kov}},\ and\ \bibinfo {author} {\bibfnamefont {I.~Y.}\ \bibnamefont
			{Dzyaloshinskii}},\ }\href@noop {} {\emph {\bibinfo {title} {Quantum Field
				Theoretical Methods in Statistical Physics}}}\ (\bibinfo  {publisher}
	{Pergamon Press Ltd., Oxford},\ \bibinfo {year} {1965})\BibitemShut {NoStop}%
	\bibitem [{\citenamefont {Fausti}\ \emph {et~al.}(2011)\citenamefont {Fausti},
		\citenamefont {Tobey}, \citenamefont {Dean}, \citenamefont {Kaiser},
		\citenamefont {Dienst}, \citenamefont {Hoffmann}, \citenamefont {Pyon},
		\citenamefont {Takayama}, \citenamefont {Takagi},\ and\ \citenamefont
		{Cavalleri}}]{fausti_light-induced_2011}%
	\BibitemOpen
	\bibfield  {author} {\bibinfo {author} {\bibfnamefont {D.}~\bibnamefont
			{Fausti}}, \bibinfo {author} {\bibfnamefont {R.~I.}\ \bibnamefont {Tobey}},
		\bibinfo {author} {\bibfnamefont {N.}~\bibnamefont {Dean}}, \bibinfo {author}
		{\bibfnamefont {S.}~\bibnamefont {Kaiser}}, \bibinfo {author} {\bibfnamefont
			{A.}~\bibnamefont {Dienst}}, \bibinfo {author} {\bibfnamefont {M.~C.}\
			\bibnamefont {Hoffmann}}, \bibinfo {author} {\bibfnamefont {S.}~\bibnamefont
			{Pyon}}, \bibinfo {author} {\bibfnamefont {T.}~\bibnamefont {Takayama}},
		\bibinfo {author} {\bibfnamefont {H.}~\bibnamefont {Takagi}},\ and\ \bibinfo
		{author} {\bibfnamefont {A.}~\bibnamefont {Cavalleri}},\ }\bibfield  {title}
	{\bibinfo {title} {Light-{Induced} {Superconductivity} in a
			{Stripe}-{Ordered} {Cuprate}},\ }\href
	{https://doi.org/10.1126/science.1197294} {\bibfield  {journal} {\bibinfo
			{journal} {Science}\ }\textbf {\bibinfo {volume} {331}},\ \bibinfo {pages}
		{189} (\bibinfo {year} {2011})}\BibitemShut {NoStop}%
	\bibitem [{\citenamefont {Nicoletti}\ \emph {et~al.}(2014)\citenamefont
		{Nicoletti}, \citenamefont {Casandruc}, \citenamefont {Laplace},
		\citenamefont {Khanna}, \citenamefont {Hunt}, \citenamefont {Kaiser},
		\citenamefont {Dhesi}, \citenamefont {Gu}, \citenamefont {Hill},\ and\
		\citenamefont {Cavalleri}}]{nicoletti_optically_2014}%
	\BibitemOpen
	\bibfield  {author} {\bibinfo {author} {\bibfnamefont {D.}~\bibnamefont
			{Nicoletti}}, \bibinfo {author} {\bibfnamefont {E.}~\bibnamefont
			{Casandruc}}, \bibinfo {author} {\bibfnamefont {Y.}~\bibnamefont {Laplace}},
		\bibinfo {author} {\bibfnamefont {V.}~\bibnamefont {Khanna}}, \bibinfo
		{author} {\bibfnamefont {C.~R.}\ \bibnamefont {Hunt}}, \bibinfo {author}
		{\bibfnamefont {S.}~\bibnamefont {Kaiser}}, \bibinfo {author} {\bibfnamefont
			{S.~S.}\ \bibnamefont {Dhesi}}, \bibinfo {author} {\bibfnamefont {G.~D.}\
			\bibnamefont {Gu}}, \bibinfo {author} {\bibfnamefont {J.~P.}\ \bibnamefont
			{Hill}},\ and\ \bibinfo {author} {\bibfnamefont {A.}~\bibnamefont
			{Cavalleri}},\ }\href {https://doi.org/10.1103/PhysRevB.90.100503} {\bibfield
		{journal} {\bibinfo  {journal} {Physical Review B}\ }\textbf {\bibinfo
			{volume} {90}},\ \bibinfo {pages} {100503(R)} (\bibinfo {year}
		{2014})}\BibitemShut {NoStop}%
	\bibitem [{\citenamefont {Cremin}\ \emph {et~al.}(2019)\citenamefont {Cremin},
		\citenamefont {Zhang}, \citenamefont {Homes}, \citenamefont {Gu},
		\citenamefont {Sun}, \citenamefont {Fogler}, \citenamefont {Millis},
		\citenamefont {Basov},\ and\ \citenamefont
		{Averitt}}]{cremin_photoenhanced_2019}%
	\BibitemOpen
	\bibfield  {author} {\bibinfo {author} {\bibfnamefont {K.~A.}\ \bibnamefont
			{Cremin}}, \bibinfo {author} {\bibfnamefont {J.}~\bibnamefont {Zhang}},
		\bibinfo {author} {\bibfnamefont {C.~C.}\ \bibnamefont {Homes}}, \bibinfo
		{author} {\bibfnamefont {G.~D.}\ \bibnamefont {Gu}}, \bibinfo {author}
		{\bibfnamefont {Z.}~\bibnamefont {Sun}}, \bibinfo {author} {\bibfnamefont
			{M.~M.}\ \bibnamefont {Fogler}}, \bibinfo {author} {\bibfnamefont {A.~J.}\
			\bibnamefont {Millis}}, \bibinfo {author} {\bibfnamefont {D.~N.}\
			\bibnamefont {Basov}},\ and\ \bibinfo {author} {\bibfnamefont {R.~D.}\
			\bibnamefont {Averitt}},\ }\bibfield  {title} {\bibinfo {title}
		{Photoenhanced metastable c-axis electrodynamics in stripe-ordered cuprate
			{La} $_{\textrm{1.885}}$ {Ba} $_{\textrm{0.115}}$ {CuO} $_{\textrm{4}}$},\
	}\href {https://doi.org/10.1073/pnas.1908368116} {\bibfield  {journal}
		{\bibinfo  {journal} {Proceedings of the National Academy of Sciences}\
		}\textbf {\bibinfo {volume} {116}},\ \bibinfo {pages} {19875} (\bibinfo
		{year} {2019})}\BibitemShut {NoStop}%
	\bibitem [{\citenamefont {Rajasekaran}\ \emph {et~al.}(2018)\citenamefont
		{Rajasekaran}, \citenamefont {Okamoto}, \citenamefont {Mathey}, \citenamefont
		{Fechner}, \citenamefont {Thampy}, \citenamefont {Gu},\ and\ \citenamefont
		{Cavalleri}}]{rajasekaran_probing_2018}%
	\BibitemOpen
	\bibfield  {author} {\bibinfo {author} {\bibfnamefont {S.}~\bibnamefont
			{Rajasekaran}}, \bibinfo {author} {\bibfnamefont {J.}~\bibnamefont
			{Okamoto}}, \bibinfo {author} {\bibfnamefont {L.}~\bibnamefont {Mathey}},
		\bibinfo {author} {\bibfnamefont {M.}~\bibnamefont {Fechner}}, \bibinfo
		{author} {\bibfnamefont {V.}~\bibnamefont {Thampy}}, \bibinfo {author}
		{\bibfnamefont {G.~D.}\ \bibnamefont {Gu}},\ and\ \bibinfo {author}
		{\bibfnamefont {A.}~\bibnamefont {Cavalleri}},\ }\bibfield  {title} {\bibinfo
		{title} {Probing optically silent superfluid stripes in cuprates},\ }\href
	{https://doi.org/10.1126/science.aan3438} {\bibfield  {journal} {\bibinfo
			{journal} {Science}\ }\textbf {\bibinfo {volume} {359}},\ \bibinfo {pages}
		{575} (\bibinfo {year} {2018})}\BibitemShut {NoStop}%
	\bibitem [{\citenamefont {Suzuki}\ \emph {et~al.}(2019)\citenamefont {Suzuki},
		\citenamefont {Someya}, \citenamefont {Hashimoto}, \citenamefont {Michimae},
		\citenamefont {Watanabe}, \citenamefont {Fujisawa}, \citenamefont {Kanai},
		\citenamefont {Ishii}, \citenamefont {Itatani}, \citenamefont {Kasahara},
		\citenamefont {Matsuda}, \citenamefont {Shibauchi}, \citenamefont {Okazaki},\
		and\ \citenamefont {Shin}}]{suzuki_photoinduced_2019}%
	\BibitemOpen
	\bibfield  {author} {\bibinfo {author} {\bibfnamefont {T.}~\bibnamefont
			{Suzuki}}, \bibinfo {author} {\bibfnamefont {T.}~\bibnamefont {Someya}},
		\bibinfo {author} {\bibfnamefont {T.}~\bibnamefont {Hashimoto}}, \bibinfo
		{author} {\bibfnamefont {S.}~\bibnamefont {Michimae}}, \bibinfo {author}
		{\bibfnamefont {M.}~\bibnamefont {Watanabe}}, \bibinfo {author}
		{\bibfnamefont {M.}~\bibnamefont {Fujisawa}}, \bibinfo {author}
		{\bibfnamefont {T.}~\bibnamefont {Kanai}}, \bibinfo {author} {\bibfnamefont
			{N.}~\bibnamefont {Ishii}}, \bibinfo {author} {\bibfnamefont
			{J.}~\bibnamefont {Itatani}}, \bibinfo {author} {\bibfnamefont
			{S.}~\bibnamefont {Kasahara}}, \bibinfo {author} {\bibfnamefont
			{Y.}~\bibnamefont {Matsuda}}, \bibinfo {author} {\bibfnamefont
			{T.}~\bibnamefont {Shibauchi}}, \bibinfo {author} {\bibfnamefont
			{K.}~\bibnamefont {Okazaki}},\ and\ \bibinfo {author} {\bibfnamefont
			{S.}~\bibnamefont {Shin}},\ }\bibfield  {title} {\bibinfo {title}
		{Photoinduced possible superconducting state with long-lived disproportionate
			band filling in {FeSe}},\ }\href {https://doi.org/10.1038/s42005-019-0219-4}
	{\bibfield  {journal} {\bibinfo  {journal} {Communications Physics}\ }\textbf
		{\bibinfo {volume} {2}},\ \bibinfo {pages} {115} (\bibinfo {year}
		{2019})}\BibitemShut {NoStop}%
	\bibitem [{\citenamefont {Mitrano}\ \emph {et~al.}(2016)\citenamefont
		{Mitrano}, \citenamefont {Cantaluppi}, \citenamefont {Nicoletti},
		\citenamefont {Kaiser}, \citenamefont {Perucchi}, \citenamefont {Lupi},
		\citenamefont {Di~Pietro}, \citenamefont {Pontiroli}, \citenamefont {Riccò},
		\citenamefont {Clark}, \citenamefont {Jaksch},\ and\ \citenamefont
		{Cavalleri}}]{mitrano_possible_2016}%
	\BibitemOpen
	\bibfield  {author} {\bibinfo {author} {\bibfnamefont {M.}~\bibnamefont
			{Mitrano}}, \bibinfo {author} {\bibfnamefont {A.}~\bibnamefont {Cantaluppi}},
		\bibinfo {author} {\bibfnamefont {D.}~\bibnamefont {Nicoletti}}, \bibinfo
		{author} {\bibfnamefont {S.}~\bibnamefont {Kaiser}}, \bibinfo {author}
		{\bibfnamefont {A.}~\bibnamefont {Perucchi}}, \bibinfo {author}
		{\bibfnamefont {S.}~\bibnamefont {Lupi}}, \bibinfo {author} {\bibfnamefont
			{P.}~\bibnamefont {Di~Pietro}}, \bibinfo {author} {\bibfnamefont
			{D.}~\bibnamefont {Pontiroli}}, \bibinfo {author} {\bibfnamefont
			{M.}~\bibnamefont {Riccò}}, \bibinfo {author} {\bibfnamefont {S.~R.}\
			\bibnamefont {Clark}}, \bibinfo {author} {\bibfnamefont {D.}~\bibnamefont
			{Jaksch}},\ and\ \bibinfo {author} {\bibfnamefont {A.}~\bibnamefont
			{Cavalleri}},\ }\bibfield  {title} {\bibinfo {title} {Possible light-induced
			superconductivity in {K3C60} at high temperature},\ }\href
	{https://doi.org/10.1038/nature16522} {\bibfield  {journal} {\bibinfo
			{journal} {Nature}\ }\textbf {\bibinfo {volume} {530}},\ \bibinfo {pages}
		{461} (\bibinfo {year} {2016})}\BibitemShut {NoStop}%
	\bibitem [{\citenamefont {Budden}\ \emph {et~al.}(2021)\citenamefont {Budden},
		\citenamefont {Gebert}, \citenamefont {Buzzi}, \citenamefont {Jotzu},
		\citenamefont {Wang}, \citenamefont {Matsuyama}, \citenamefont {Meier},
		\citenamefont {Laplace}, \citenamefont {Pontiroli}, \citenamefont {Riccò},
		\citenamefont {Schlawin}, \citenamefont {Jaksch},\ and\ \citenamefont
		{Cavalleri}}]{budden_evidence_2021}%
	\BibitemOpen
	\bibfield  {author} {\bibinfo {author} {\bibfnamefont {M.}~\bibnamefont
			{Budden}}, \bibinfo {author} {\bibfnamefont {T.}~\bibnamefont {Gebert}},
		\bibinfo {author} {\bibfnamefont {M.}~\bibnamefont {Buzzi}}, \bibinfo
		{author} {\bibfnamefont {G.}~\bibnamefont {Jotzu}}, \bibinfo {author}
		{\bibfnamefont {E.}~\bibnamefont {Wang}}, \bibinfo {author} {\bibfnamefont
			{T.}~\bibnamefont {Matsuyama}}, \bibinfo {author} {\bibfnamefont
			{G.}~\bibnamefont {Meier}}, \bibinfo {author} {\bibfnamefont
			{Y.}~\bibnamefont {Laplace}}, \bibinfo {author} {\bibfnamefont
			{D.}~\bibnamefont {Pontiroli}}, \bibinfo {author} {\bibfnamefont
			{M.}~\bibnamefont {Riccò}}, \bibinfo {author} {\bibfnamefont
			{F.}~\bibnamefont {Schlawin}}, \bibinfo {author} {\bibfnamefont
			{D.}~\bibnamefont {Jaksch}},\ and\ \bibinfo {author} {\bibfnamefont
			{A.}~\bibnamefont {Cavalleri}},\ }\bibfield  {title} {\bibinfo {title}
		{Evidence for metastable photo-induced superconductivity in {K3C60}},\ }\href
	{https://doi.org/10.1038/s41567-020-01148-1} {\bibfield  {journal} {\bibinfo
			{journal} {Nature Physics}\ }\textbf {\bibinfo {volume} {17}},\ \bibinfo
		{pages} {611} (\bibinfo {year} {2021})}\BibitemShut {NoStop}%
	\bibitem [{\citenamefont {Buzzi}\ \emph {et~al.}(2020)\citenamefont {Buzzi},
		\citenamefont {Nicoletti}, \citenamefont {Fechner}, \citenamefont
		{Tancogne-Dejean}, \citenamefont {Sentef}, \citenamefont {Georges},
		\citenamefont {Biesner}, \citenamefont {Uykur}, \citenamefont {Dressel},
		\citenamefont {Henderson}, \citenamefont {Siegrist}, \citenamefont
		{Schlueter}, \citenamefont {Miyagawa}, \citenamefont {Kanoda}, \citenamefont
		{Nam}, \citenamefont {Ardavan}, \citenamefont {Coulthard}, \citenamefont
		{Tindall}, \citenamefont {Schlawin}, \citenamefont {Jaksch},\ and\
		\citenamefont {Cavalleri}}]{buzzi_photomolecular_2020}%
	\BibitemOpen
	\bibfield  {author} {\bibinfo {author} {\bibfnamefont {M.}~\bibnamefont
			{Buzzi}}, \bibinfo {author} {\bibfnamefont {D.}~\bibnamefont {Nicoletti}},
		\bibinfo {author} {\bibfnamefont {M.}~\bibnamefont {Fechner}}, \bibinfo
		{author} {\bibfnamefont {N.}~\bibnamefont {Tancogne-Dejean}}, \bibinfo
		{author} {\bibfnamefont {M.}~\bibnamefont {Sentef}}, \bibinfo {author}
		{\bibfnamefont {A.}~\bibnamefont {Georges}}, \bibinfo {author} {\bibfnamefont
			{T.}~\bibnamefont {Biesner}}, \bibinfo {author} {\bibfnamefont
			{E.}~\bibnamefont {Uykur}}, \bibinfo {author} {\bibfnamefont
			{M.}~\bibnamefont {Dressel}}, \bibinfo {author} {\bibfnamefont
			{A.}~\bibnamefont {Henderson}}, \bibinfo {author} {\bibfnamefont
			{T.}~\bibnamefont {Siegrist}}, \bibinfo {author} {\bibfnamefont
			{J.}~\bibnamefont {Schlueter}}, \bibinfo {author} {\bibfnamefont
			{K.}~\bibnamefont {Miyagawa}}, \bibinfo {author} {\bibfnamefont
			{K.}~\bibnamefont {Kanoda}}, \bibinfo {author} {\bibfnamefont {M.-S.}\
			\bibnamefont {Nam}}, \bibinfo {author} {\bibfnamefont {A.}~\bibnamefont
			{Ardavan}}, \bibinfo {author} {\bibfnamefont {J.}~\bibnamefont {Coulthard}},
		\bibinfo {author} {\bibfnamefont {J.}~\bibnamefont {Tindall}}, \bibinfo
		{author} {\bibfnamefont {F.}~\bibnamefont {Schlawin}}, \bibinfo {author}
		{\bibfnamefont {D.}~\bibnamefont {Jaksch}},\ and\ \bibinfo {author}
		{\bibfnamefont {A.}~\bibnamefont {Cavalleri}},\ }\bibfield  {title} {\bibinfo
		{title} {Photomolecular {High}-{Temperature} {Superconductivity}},\ }\href
	{https://doi.org/10.1103/PhysRevX.10.031028} {\bibfield  {journal} {\bibinfo
			{journal} {Physical Review X}\ }\textbf {\bibinfo {volume} {10}},\ \bibinfo
		{pages} {031028} (\bibinfo {year} {2020})}\BibitemShut {NoStop}%
	\bibitem [{\citenamefont {Bloch}\ \emph {et~al.}(2022)\citenamefont {Bloch},
		\citenamefont {Cavalleri}, \citenamefont {Galitski}, \citenamefont {Hafezi},\
		and\ \citenamefont {Rubio}}]{bloch_strongly_2022}%
	\BibitemOpen
	\bibfield  {author} {\bibinfo {author} {\bibfnamefont {J.}~\bibnamefont
			{Bloch}}, \bibinfo {author} {\bibfnamefont {A.}~\bibnamefont {Cavalleri}},
		\bibinfo {author} {\bibfnamefont {V.}~\bibnamefont {Galitski}}, \bibinfo
		{author} {\bibfnamefont {M.}~\bibnamefont {Hafezi}},\ and\ \bibinfo {author}
		{\bibfnamefont {A.}~\bibnamefont {Rubio}},\ }\bibfield  {title} {\bibinfo
		{title} {Strongly correlated electron–photon systems},\ }\href
	{https://doi.org/10.1038/s41586-022-04726-w} {\bibfield  {journal} {\bibinfo
			{journal} {Nature}\ }\textbf {\bibinfo {volume} {606}},\ \bibinfo {pages}
		{41} (\bibinfo {year} {2022})}\BibitemShut {NoStop}%
	\bibitem [{\citenamefont {Sentef}\ \emph {et~al.}(2016)\citenamefont {Sentef},
		\citenamefont {Kemper}, \citenamefont {Georges},\ and\ \citenamefont
		{Kollath}}]{sentef_theory_2016}%
	\BibitemOpen
	\bibfield  {author} {\bibinfo {author} {\bibfnamefont {M.~A.}\ \bibnamefont
			{Sentef}}, \bibinfo {author} {\bibfnamefont {A.~F.}\ \bibnamefont {Kemper}},
		\bibinfo {author} {\bibfnamefont {A.}~\bibnamefont {Georges}},\ and\ \bibinfo
		{author} {\bibfnamefont {C.}~\bibnamefont {Kollath}},\ }\bibfield  {title}
	{\bibinfo {title} {Theory of light-enhanced phonon-mediated
			superconductivity},\ }\href {https://doi.org/10.1103/PhysRevB.93.144506}
	{\bibfield  {journal} {\bibinfo  {journal} {Phys. Rev. B}\ }\textbf {\bibinfo
			{volume} {93}},\ \bibinfo {pages} {144506} (\bibinfo {year}
		{2016})}\BibitemShut {NoStop}%
	\bibitem [{\citenamefont {Tindall}\ \emph {et~al.}(2019)\citenamefont
		{Tindall}, \citenamefont {Buča}, \citenamefont {Coulthard},\ and\
		\citenamefont {Jaksch}}]{tindall_heating-induced_2019}%
	\BibitemOpen
	\bibfield  {author} {\bibinfo {author} {\bibfnamefont {J.}~\bibnamefont
			{Tindall}}, \bibinfo {author} {\bibfnamefont {B.}~\bibnamefont {Buča}},
		\bibinfo {author} {\bibfnamefont {J.}~\bibnamefont {Coulthard}},\ and\
		\bibinfo {author} {\bibfnamefont {D.}~\bibnamefont {Jaksch}},\ }\bibfield
	{title} {\bibinfo {title} {Heating-{Induced} {Long}-{Range} $\eta$ {Pairing}
			in the {Hubbard} {Model}},\ }\href
	{https://doi.org/10.1103/PhysRevLett.123.030603} {\bibfield  {journal}
		{\bibinfo  {journal} {Physical Review Letters}\ }\textbf {\bibinfo {volume}
			{123}},\ \bibinfo {pages} {030603} (\bibinfo {year} {2019})}\BibitemShut
	{NoStop}%
	\bibitem [{\citenamefont {Kaneko}\ \emph {et~al.}(2019)\citenamefont {Kaneko},
		\citenamefont {Shirakawa}, \citenamefont {Sorella},\ and\ \citenamefont
		{Yunoki}}]{kaneko_photoinduced_2019}%
	\BibitemOpen
	\bibfield  {author} {\bibinfo {author} {\bibfnamefont {T.}~\bibnamefont
			{Kaneko}}, \bibinfo {author} {\bibfnamefont {T.}~\bibnamefont {Shirakawa}},
		\bibinfo {author} {\bibfnamefont {S.}~\bibnamefont {Sorella}},\ and\ \bibinfo
		{author} {\bibfnamefont {S.}~\bibnamefont {Yunoki}},\ }\bibfield  {title}
	{\bibinfo {title} {Photoinduced $\eta$ {Pairing} in the {Hubbard} {Model}},\
	}\href {https://doi.org/10.1103/PhysRevLett.122.077002} {\bibfield  {journal}
		{\bibinfo  {journal} {Physical Review Letters}\ }\textbf {\bibinfo {volume}
			{122}},\ \bibinfo {pages} {077002} (\bibinfo {year} {2019})}\BibitemShut
	{NoStop}%
	\bibitem [{\citenamefont {Cook}\ and\ \citenamefont
		{Clark}(2020)}]{cook_controllable_2020}%
	\BibitemOpen
	\bibfield  {author} {\bibinfo {author} {\bibfnamefont {M.~W.}\ \bibnamefont
			{Cook}}\ and\ \bibinfo {author} {\bibfnamefont {S.~R.}\ \bibnamefont
			{Clark}},\ }\bibfield  {title} {\bibinfo {title} {Controllable finite-momenta
			dynamical quasicondensation in the periodically driven one-dimensional
			{Fermi}-{Hubbard} model},\ }\href
	{https://doi.org/10.1103/PhysRevA.101.033604} {\bibfield  {journal} {\bibinfo
			{journal} {Physical Review A}\ }\textbf {\bibinfo {volume} {101}},\ \bibinfo
		{pages} {033604} (\bibinfo {year} {2020})}\BibitemShut {NoStop}%
	\bibitem [{\citenamefont {Paeckel}\ \emph {et~al.}(2020)\citenamefont
		{Paeckel}, \citenamefont {Fauseweh}, \citenamefont {Osterkorn}, \citenamefont
		{K\"ohler}, \citenamefont {Manske},\ and\ \citenamefont
		{Manmana}}]{paeckel_detecting_2020}%
	\BibitemOpen
	\bibfield  {author} {\bibinfo {author} {\bibfnamefont {S.}~\bibnamefont
			{Paeckel}}, \bibinfo {author} {\bibfnamefont {B.}~\bibnamefont {Fauseweh}},
		\bibinfo {author} {\bibfnamefont {A.}~\bibnamefont {Osterkorn}}, \bibinfo
		{author} {\bibfnamefont {T.}~\bibnamefont {K\"ohler}}, \bibinfo {author}
		{\bibfnamefont {D.}~\bibnamefont {Manske}},\ and\ \bibinfo {author}
		{\bibfnamefont {S.~R.}\ \bibnamefont {Manmana}},\ }\bibfield  {title}
	{\bibinfo {title} {Detecting superconductivity out of equilibrium},\ }\href
	{https://doi.org/10.1103/PhysRevB.101.180507} {\bibfield  {journal} {\bibinfo
			{journal} {Physical Review B}\ }\textbf {\bibinfo {volume} {101}},\ \bibinfo
		{pages} {180507(R)} (\bibinfo {year} {2020})}\BibitemShut {NoStop}%
	\bibitem [{\citenamefont {Buzzi}\ \emph {et~al.}(2021)\citenamefont {Buzzi},
		\citenamefont {Jotzu}, \citenamefont {Cavalleri}, \citenamefont {Cirac},
		\citenamefont {Demler}, \citenamefont {Halperin}, \citenamefont {Lukin},
		\citenamefont {Shi}, \citenamefont {Wang},\ and\ \citenamefont
		{Podolsky}}]{buzzi_higgs_2021}%
	\BibitemOpen
	\bibfield  {author} {\bibinfo {author} {\bibfnamefont {M.}~\bibnamefont
			{Buzzi}}, \bibinfo {author} {\bibfnamefont {G.}~\bibnamefont {Jotzu}},
		\bibinfo {author} {\bibfnamefont {A.}~\bibnamefont {Cavalleri}}, \bibinfo
		{author} {\bibfnamefont {J.~I.}\ \bibnamefont {Cirac}}, \bibinfo {author}
		{\bibfnamefont {E.~A.}\ \bibnamefont {Demler}}, \bibinfo {author}
		{\bibfnamefont {B.~I.}\ \bibnamefont {Halperin}}, \bibinfo {author}
		{\bibfnamefont {M.~D.}\ \bibnamefont {Lukin}}, \bibinfo {author}
		{\bibfnamefont {T.}~\bibnamefont {Shi}}, \bibinfo {author} {\bibfnamefont
			{Y.}~\bibnamefont {Wang}},\ and\ \bibinfo {author} {\bibfnamefont
			{D.}~\bibnamefont {Podolsky}},\ }\bibfield  {title} {\bibinfo {title}
		{Higgs-mediated optical amplification in a nonequilibrium superconductor},\
	}\href {https://doi.org/10.1103/PhysRevX.11.011055} {\bibfield  {journal}
		{\bibinfo  {journal} {Phys. Rev. X}\ }\textbf {\bibinfo {volume} {11}},\
		\bibinfo {pages} {011055} (\bibinfo {year} {2021})}\BibitemShut {NoStop}%
	\bibitem [{\citenamefont {Dolgirev}\ \emph {et~al.}(2022)\citenamefont
		{Dolgirev}, \citenamefont {Zong}, \citenamefont {Michael}, \citenamefont
		{Curtis}, \citenamefont {Podolsky}, \citenamefont {Cavalleri},\ and\
		\citenamefont {Demler}}]{dolgirev_periodic_2022}%
	\BibitemOpen
	\bibfield  {author} {\bibinfo {author} {\bibfnamefont {P.~E.}\ \bibnamefont
			{Dolgirev}}, \bibinfo {author} {\bibfnamefont {A.}~\bibnamefont {Zong}},
		\bibinfo {author} {\bibfnamefont {M.~H.}\ \bibnamefont {Michael}}, \bibinfo
		{author} {\bibfnamefont {J.~B.}\ \bibnamefont {Curtis}}, \bibinfo {author}
		{\bibfnamefont {D.}~\bibnamefont {Podolsky}}, \bibinfo {author}
		{\bibfnamefont {A.}~\bibnamefont {Cavalleri}},\ and\ \bibinfo {author}
		{\bibfnamefont {E.}~\bibnamefont {Demler}},\ }\bibfield  {title} {\bibinfo
		{title} {Periodic dynamics in superconductors induced by an impulsive optical
			quench},\ }\href {https://doi.org/10.1038/s42005-022-01007-w} {\bibfield
		{journal} {\bibinfo  {journal} {Communications Physics}\ }\textbf {\bibinfo
			{volume} {5}},\ \bibinfo {pages} {234} (\bibinfo {year} {2022})}\BibitemShut
	{NoStop}%
	\bibitem [{\citenamefont {Hubbard}(1963)}]{hubbard_electron_1963}%
	\BibitemOpen
	\bibfield  {author} {\bibinfo {author} {\bibfnamefont {J.}~\bibnamefont
			{Hubbard}},\ }\bibfield  {title} {\bibinfo {title} {Electron correlations in
			narrow energy bands},\ }\href {https://doi.org/10.1098/rspa.1963.0204}
	{\bibfield  {journal} {\bibinfo  {journal} {Proceedings of the Royal Society
				of London. Series A. Mathematical and Physical Sciences}\ }\textbf {\bibinfo
			{volume} {276}},\ \bibinfo {pages} {238} (\bibinfo {year}
		{1963})}\BibitemShut {NoStop}%
	\bibitem [{\citenamefont {Hubbard}(1964)}]{hubbard_electron_1964}%
	\BibitemOpen
	\bibfield  {author} {\bibinfo {author} {\bibfnamefont {J.}~\bibnamefont
			{Hubbard}},\ }\bibfield  {title} {\bibinfo {title} {Electron correlations in
			narrow energy bands {III}. {An} improved solution},\ }\href
	{https://doi.org/10.1098/rspa.1964.0190} {\bibfield  {journal} {\bibinfo
			{journal} {Proceedings of the Royal Society of London. Series A. Mathematical
				and Physical Sciences}\ }\textbf {\bibinfo {volume} {281}},\ \bibinfo {pages}
		{401} (\bibinfo {year} {1964})}\BibitemShut {NoStop}%
	\bibitem [{\citenamefont {Gutzwiller}(1963)}]{gutzwiller_effect_1963}%
	\BibitemOpen
	\bibfield  {author} {\bibinfo {author} {\bibfnamefont {M.~C.}\ \bibnamefont
			{Gutzwiller}},\ }\bibfield  {title} {\bibinfo {title} {Effect of
			{Correlation} on the {Ferromagnetism} of {Transition} {Metals}},\ }\href
	{https://doi.org/10.1103/PhysRevLett.10.159} {\bibfield  {journal} {\bibinfo
			{journal} {Physical Review Letters}\ }\textbf {\bibinfo {volume} {10}},\
		\bibinfo {pages} {159} (\bibinfo {year} {1963})}\BibitemShut {NoStop}%
	\bibitem [{\citenamefont {Kanamori}(1963)}]{kanamori_electron_1963}%
	\BibitemOpen
	\bibfield  {author} {\bibinfo {author} {\bibfnamefont {J.}~\bibnamefont
			{Kanamori}},\ }\bibfield  {title} {\bibinfo {title} {Electron {Correlation}
			and {Ferromagnetism} of {Transition} {Metals}},\ }\href
	{https://doi.org/10.1143/PTP.30.275} {\bibfield  {journal} {\bibinfo
			{journal} {Progress of Theoretical Physics}\ }\textbf {\bibinfo {volume}
			{30}},\ \bibinfo {pages} {275} (\bibinfo {year} {1963})}\BibitemShut
	{NoStop}%
	\bibitem [{\citenamefont {Qin}\ \emph {et~al.}(2022)\citenamefont {Qin},
		\citenamefont {Schäfer}, \citenamefont {Andergassen}, \citenamefont
		{Corboz},\ and\ \citenamefont {Gull}}]{qin_hubbard_2022}%
	\BibitemOpen
	\bibfield  {author} {\bibinfo {author} {\bibfnamefont {M.}~\bibnamefont
			{Qin}}, \bibinfo {author} {\bibfnamefont {T.}~\bibnamefont {Schäfer}},
		\bibinfo {author} {\bibfnamefont {S.}~\bibnamefont {Andergassen}}, \bibinfo
		{author} {\bibfnamefont {P.}~\bibnamefont {Corboz}},\ and\ \bibinfo {author}
		{\bibfnamefont {E.}~\bibnamefont {Gull}},\ }\bibfield  {title} {\bibinfo
		{title} {The {Hubbard} {Model}: {A} {Computational} {Perspective}},\ }\href
	{https://doi.org/10.1146/annurev-conmatphys-090921-033948} {\bibfield
		{journal} {\bibinfo  {journal} {Annual Review of Condensed Matter Physics}\
		}\textbf {\bibinfo {volume} {13}},\ \bibinfo {pages} {275} (\bibinfo {year}
		{2022})}\BibitemShut {NoStop}%
	\bibitem [{\citenamefont {Yang}(1989)}]{yang__1989}%
	\BibitemOpen
	\bibfield  {author} {\bibinfo {author} {\bibfnamefont {C.~N.}\ \bibnamefont
			{Yang}},\ }\bibfield  {title} {\bibinfo {title} {$\eta$ pairing and
			off-diagonal long-range order in a {Hubbard} model},\ }\href
	{https://doi.org/10.1103/PhysRevLett.63.2144} {\bibfield  {journal} {\bibinfo
			{journal} {Physical Review Letters}\ }\textbf {\bibinfo {volume} {63}},\
		\bibinfo {pages} {2144} (\bibinfo {year} {1989})}\BibitemShut {NoStop}%
	\bibitem [{\citenamefont {Yang}(1962)}]{yang_concept_1962}%
	\BibitemOpen
	\bibfield  {author} {\bibinfo {author} {\bibfnamefont {C.~N.}\ \bibnamefont
			{Yang}},\ }\bibfield  {title} {\bibinfo {title} {Concept of {Off}-{Diagonal}
			{Long}-{Range} {Order} and the {Quantum} {Phases} of {Liquid} {He} and of
			{Superconductors}},\ }\href {https://doi.org/10.1103/RevModPhys.34.694}
	{\bibfield  {journal} {\bibinfo  {journal} {Reviews of Modern Physics}\
		}\textbf {\bibinfo {volume} {34}},\ \bibinfo {pages} {694} (\bibinfo {year}
		{1962})}\BibitemShut {NoStop}%
	\bibitem [{\citenamefont {Rigol}\ and\ \citenamefont
		{Muramatsu}(2004)}]{rigol_emergence_2004}%
	\BibitemOpen
	\bibfield  {author} {\bibinfo {author} {\bibfnamefont {M.}~\bibnamefont
			{Rigol}}\ and\ \bibinfo {author} {\bibfnamefont {A.}~\bibnamefont
			{Muramatsu}},\ }\bibfield  {title} {\bibinfo {title} {Emergence of
			{Quasicondensates} of {Hard}-{Core} {Bosons} at {Finite} {Momentum}},\ }\href
	{https://doi.org/10.1103/PhysRevLett.93.230404} {\bibfield  {journal}
		{\bibinfo  {journal} {Physical Review Letters}\ }\textbf {\bibinfo {volume}
			{93}},\ \bibinfo {pages} {230404} (\bibinfo {year} {2004})}\BibitemShut
	{NoStop}%
	\bibitem [{\citenamefont {Rodriguez}\ \emph {et~al.}(2006)\citenamefont
		{Rodriguez}, \citenamefont {Manmana}, \citenamefont {Rigol}, \citenamefont
		{Noack},\ and\ \citenamefont {Muramatsu}}]{rodriguez_coherent_2006}%
	\BibitemOpen
	\bibfield  {author} {\bibinfo {author} {\bibfnamefont {K.}~\bibnamefont
			{Rodriguez}}, \bibinfo {author} {\bibfnamefont {S.~R.}\ \bibnamefont
			{Manmana}}, \bibinfo {author} {\bibfnamefont {M.}~\bibnamefont {Rigol}},
		\bibinfo {author} {\bibfnamefont {R.~M.}\ \bibnamefont {Noack}},\ and\
		\bibinfo {author} {\bibfnamefont {A.}~\bibnamefont {Muramatsu}},\ }\bibfield
	{title} {\bibinfo {title} {Coherent matter waves emerging from
			{Mott}-insulators},\ }\href {https://doi.org/10.1088/1367-2630/8/8/169}
	{\bibfield  {journal} {\bibinfo  {journal} {New Journal of Physics}\ }\textbf
		{\bibinfo {volume} {8}},\ \bibinfo {pages} {169} (\bibinfo {year}
		{2006})}\BibitemShut {NoStop}%
	\bibitem [{\citenamefont {Vidmar}\ \emph {et~al.}(2015)\citenamefont {Vidmar},
		\citenamefont {Ronzheimer}, \citenamefont {Schreiber}, \citenamefont {Braun},
		\citenamefont {Hodgman}, \citenamefont {Langer}, \citenamefont
		{Heidrich-Meisner}, \citenamefont {Bloch},\ and\ \citenamefont
		{Schneider}}]{vidmar_dynamical_2015}%
	\BibitemOpen
	\bibfield  {author} {\bibinfo {author} {\bibfnamefont {L.}~\bibnamefont
			{Vidmar}}, \bibinfo {author} {\bibfnamefont {J.~P.}\ \bibnamefont
			{Ronzheimer}}, \bibinfo {author} {\bibfnamefont {M.}~\bibnamefont
			{Schreiber}}, \bibinfo {author} {\bibfnamefont {S.}~\bibnamefont {Braun}},
		\bibinfo {author} {\bibfnamefont {S.~S.}\ \bibnamefont {Hodgman}}, \bibinfo
		{author} {\bibfnamefont {S.}~\bibnamefont {Langer}}, \bibinfo {author}
		{\bibfnamefont {F.}~\bibnamefont {Heidrich-Meisner}}, \bibinfo {author}
		{\bibfnamefont {I.}~\bibnamefont {Bloch}},\ and\ \bibinfo {author}
		{\bibfnamefont {U.}~\bibnamefont {Schneider}},\ }\bibfield  {title} {\bibinfo
		{title} {Dynamical quasicondensation of hard-core bosons at finite momenta},\
	}\href {https://doi.org/10.1103/PhysRevLett.115.175301} {\bibfield  {journal}
		{\bibinfo  {journal} {Phys. Rev. Lett.}\ }\textbf {\bibinfo {volume} {115}},\
		\bibinfo {pages} {175301} (\bibinfo {year} {2015})}\BibitemShut {NoStop}%
	\bibitem [{\citenamefont {Vidmar}\ \emph {et~al.}(2017)\citenamefont {Vidmar},
		\citenamefont {Iyer},\ and\ \citenamefont {Rigol}}]{vidmar_emergent_2017}%
	\BibitemOpen
	\bibfield  {author} {\bibinfo {author} {\bibfnamefont {L.}~\bibnamefont
			{Vidmar}}, \bibinfo {author} {\bibfnamefont {D.}~\bibnamefont {Iyer}},\ and\
		\bibinfo {author} {\bibfnamefont {M.}~\bibnamefont {Rigol}},\ }\bibfield
	{title} {\bibinfo {title} {Emergent {Eigenstate} {Solution} to {Quantum}
			{Dynamics} {Far} from {Equilibrium}},\ }\href
	{https://doi.org/10.1103/PhysRevX.7.021012} {\bibfield  {journal} {\bibinfo
			{journal} {Physical Review X}\ }\textbf {\bibinfo {volume} {7}},\ \bibinfo
		{pages} {021012} (\bibinfo {year} {2017})}\BibitemShut {NoStop}%
	\bibitem [{\citenamefont {Heidrich-Meisner}\ \emph {et~al.}(2008)\citenamefont
		{Heidrich-Meisner}, \citenamefont {Rigol}, \citenamefont {Muramatsu},
		\citenamefont {Feiguin},\ and\ \citenamefont
		{Dagotto}}]{heidrich-meisner_ground-state_2008}%
	\BibitemOpen
	\bibfield  {author} {\bibinfo {author} {\bibfnamefont {F.}~\bibnamefont
			{Heidrich-Meisner}}, \bibinfo {author} {\bibfnamefont {M.}~\bibnamefont
			{Rigol}}, \bibinfo {author} {\bibfnamefont {A.}~\bibnamefont {Muramatsu}},
		\bibinfo {author} {\bibfnamefont {A.~E.}\ \bibnamefont {Feiguin}},\ and\
		\bibinfo {author} {\bibfnamefont {E.}~\bibnamefont {Dagotto}},\ }\bibfield
	{title} {\bibinfo {title} {Ground-state reference systems for expanding
			correlated fermions in one dimension},\ }\href
	{https://doi.org/10.1103/PhysRevA.78.013620} {\bibfield  {journal} {\bibinfo
			{journal} {Physical Review A}\ }\textbf {\bibinfo {volume} {78}},\ \bibinfo
		{pages} {013620} (\bibinfo {year} {2008})}\BibitemShut {NoStop}%
	\bibitem [{\citenamefont {Lackner}\ \emph {et~al.}(2015)\citenamefont
		{Lackner}, \citenamefont {B\v rezinov\'a}, \citenamefont {Sato}, \citenamefont
		{Ishikawa},\ and\ \citenamefont {Burgd\"orfer}}]{lackner_propagating_2015}%
	\BibitemOpen
	\bibfield  {author} {\bibinfo {author} {\bibfnamefont {F.}~\bibnamefont
			{Lackner}}, \bibinfo {author} {\bibfnamefont {I.}~\bibnamefont
			{B\v rezinov\'a}}, \bibinfo {author} {\bibfnamefont {T.}~\bibnamefont {Sato}},
		\bibinfo {author} {\bibfnamefont {K.~L.}\ \bibnamefont {Ishikawa}},\ and\
		\bibinfo {author} {\bibfnamefont {J.}~\bibnamefont {Burgd\"orfer}},\
	}\bibfield  {title} {\bibinfo {title} {Propagating two-particle reduced
			density matrices without wave functions},\ }\href
	{https://doi.org/10.1103/PhysRevA.91.023412} {\bibfield  {journal} {\bibinfo
			{journal} {Physical Review A}\ }\textbf {\bibinfo {volume} {91}},\ \bibinfo
		{pages} {023412} (\bibinfo {year} {2015})}\BibitemShut {NoStop}%
	\bibitem [{\citenamefont {Lackner}\ \emph {et~al.}(2017)\citenamefont
		{Lackner}, \citenamefont {Březinová}, \citenamefont {Sato}, \citenamefont
		{Ishikawa},\ and\ \citenamefont {Burgd\"orfer}}]{lackner_high-harmonic_2017}%
	\BibitemOpen
	\bibfield  {author} {\bibinfo {author} {\bibfnamefont {F.}~\bibnamefont
			{Lackner}}, \bibinfo {author} {\bibfnamefont {I.}~\bibnamefont
			{Březinová}}, \bibinfo {author} {\bibfnamefont {T.}~\bibnamefont {Sato}},
		\bibinfo {author} {\bibfnamefont {K.~L.}\ \bibnamefont {Ishikawa}},\ and\
		\bibinfo {author} {\bibfnamefont {J.}~\bibnamefont {Burgd\"orfer}},\
	}\bibfield  {title} {\bibinfo {title} {High-harmonic spectra from
			time-dependent two-particle reduced-density-matrix theory},\ }\href
	{https://doi.org/10.1103/PhysRevA.95.033414} {\bibfield  {journal} {\bibinfo
			{journal} {Physical Review A}\ }\textbf {\bibinfo {volume} {95}},\ \bibinfo
		{pages} {033414} (\bibinfo {year} {2017})}\BibitemShut {NoStop}%
	\bibitem [{\citenamefont {Donsa}\ \emph {et~al.}(2023)\citenamefont {Donsa},
		\citenamefont {Lackner}, \citenamefont {Burgd\"orfer}, \citenamefont {Bonitz},
		\citenamefont {Kloss}, \citenamefont {Rubio},\ and\ \citenamefont
		{Březinová}}]{donsa_nonequilibrium_2023}%
	\BibitemOpen
	\bibfield  {author} {\bibinfo {author} {\bibfnamefont {S.}~\bibnamefont
			{Donsa}}, \bibinfo {author} {\bibfnamefont {F.}~\bibnamefont {Lackner}},
		\bibinfo {author} {\bibfnamefont {J.}~\bibnamefont {Burgd\"orfer}}, \bibinfo
		{author} {\bibfnamefont {M.}~\bibnamefont {Bonitz}}, \bibinfo {author}
		{\bibfnamefont {B.}~\bibnamefont {Kloss}}, \bibinfo {author} {\bibfnamefont
			{A.}~\bibnamefont {Rubio}},\ and\ \bibinfo {author} {\bibfnamefont
			{I.}~\bibnamefont {Březinová}},\ }\bibfield  {title} {\bibinfo {title}
		{Nonequilibrium correlation dynamics in the one-dimensional {Fermi}-{Hubbard}
			model: {A} testbed for the two-particle reduced density matrix theory},\
	}\href {https://doi.org/10.1103/PhysRevResearch.5.033022} {\bibfield
		{journal} {\bibinfo  {journal} {Physical Review Research}\ }\textbf {\bibinfo
			{volume} {5}},\ \bibinfo {pages} {033022} (\bibinfo {year}
		{2023})}\BibitemShut {NoStop}%
	\bibitem [{\citenamefont {Mermin}\ and\ \citenamefont
		{Wagner}(1966)}]{mermin_absence_1966}%
	\BibitemOpen
	\bibfield  {author} {\bibinfo {author} {\bibfnamefont {N.~D.}\ \bibnamefont
			{Mermin}}\ and\ \bibinfo {author} {\bibfnamefont {H.}~\bibnamefont
			{Wagner}},\ }\bibfield  {title} {\bibinfo {title} {Absence of
			{Ferromagnetism} or {Antiferromagnetism} in {One}- or {Two}-{Dimensional}
			{Isotropic} {Heisenberg} {Models}},\ }\href
	{https://doi.org/10.1103/PhysRevLett.17.1133} {\bibfield  {journal} {\bibinfo
			{journal} {Physical Review Letters}\ }\textbf {\bibinfo {volume} {17}},\
		\bibinfo {pages} {1133} (\bibinfo {year} {1966})}\BibitemShut {NoStop}%
	\bibitem [{\citenamefont {Hohenberg}(1967)}]{hohenberg_existence_1967}%
	\BibitemOpen
	\bibfield  {author} {\bibinfo {author} {\bibfnamefont {P.~C.}\ \bibnamefont
			{Hohenberg}},\ }\bibfield  {title} {\bibinfo {title} {Existence of
			{Long}-{Range} {Order} in {One} and {Two} {Dimensions}},\ }\href
	{https://doi.org/10.1103/PhysRev.158.383} {\bibfield  {journal} {\bibinfo
			{journal} {Physical Review}\ }\textbf {\bibinfo {volume} {158}},\ \bibinfo
		{pages} {383} (\bibinfo {year} {1967})}\BibitemShut {NoStop}%
	\bibitem [{\citenamefont {Akbari}\ \emph {et~al.}(2012)\citenamefont {Akbari},
		\citenamefont {Hashemi}, \citenamefont {Rubio}, \citenamefont {Nieminen},\
		and\ \citenamefont {van~Leeuwen}}]{akbari_challenges_2012}%
	\BibitemOpen
	\bibfield  {author} {\bibinfo {author} {\bibfnamefont {A.}~\bibnamefont
			{Akbari}}, \bibinfo {author} {\bibfnamefont {M.~J.}\ \bibnamefont {Hashemi}},
		\bibinfo {author} {\bibfnamefont {A.}~\bibnamefont {Rubio}}, \bibinfo
		{author} {\bibfnamefont {R.~M.}\ \bibnamefont {Nieminen}},\ and\ \bibinfo
		{author} {\bibfnamefont {R.}~\bibnamefont {van~Leeuwen}},\ }\bibfield
	{title} {\bibinfo {title} {Challenges in truncating the hierarchy of
			time-dependent reduced density matrices equations},\ }\href
	{https://doi.org/10.1103/PhysRevB.85.235121} {\bibfield  {journal} {\bibinfo
			{journal} {Physical Review B}\ }\textbf {\bibinfo {volume} {85}},\ \bibinfo
		{pages} {235121} (\bibinfo {year} {2012})}\BibitemShut {NoStop}%
	\bibitem [{\citenamefont {Schl\"unzen}\ \emph {et~al.}(2017)\citenamefont
		{Schl\"unzen}, \citenamefont {Joost}, \citenamefont {Heidrich-Meisner},\ and\
		\citenamefont {Bonitz}}]{schlunzen_nonequilibrium_2017}%
	\BibitemOpen
	\bibfield  {author} {\bibinfo {author} {\bibfnamefont {N.}~\bibnamefont
			{Schl\"unzen}}, \bibinfo {author} {\bibfnamefont {J.-P.}\ \bibnamefont
			{Joost}}, \bibinfo {author} {\bibfnamefont {F.}~\bibnamefont
			{Heidrich-Meisner}},\ and\ \bibinfo {author} {\bibfnamefont {M.}~\bibnamefont
			{Bonitz}},\ }\bibfield  {title} {\bibinfo {title} {Nonequilibrium dynamics in
			the one-dimensional {Fermi}-{Hubbard} model: {Comparison} of the
			nonequilibrium {Green}-functions approach and the density matrix
			renormalization group method},\ }\href
	{https://doi.org/10.1103/PhysRevB.95.165139} {\bibfield  {journal} {\bibinfo
			{journal} {Physical Review B}\ }\textbf {\bibinfo {volume} {95}},\ \bibinfo
		{pages} {165139} (\bibinfo {year} {2017})}\BibitemShut {NoStop}%
	\bibitem [{\citenamefont {Mazziotti}(1998)}]{mazziotti_approximate_1998}%
	\BibitemOpen
	\bibfield  {author} {\bibinfo {author} {\bibfnamefont {D.~A.}\ \bibnamefont
			{Mazziotti}},\ }\bibfield  {title} {\bibinfo {title} {Approximate solution
			for electron correlation through the use of {Schwinger} probes},\ }\href
	{https://doi.org/10.1016/S0009-2614(98)00470-9} {\bibfield  {journal}
		{\bibinfo  {journal} {Chemical Physics Letters}\ }\textbf {\bibinfo {volume}
			{289}},\ \bibinfo {pages} {419} (\bibinfo {year} {1998})}\BibitemShut
	{NoStop}%
	\bibitem [{\citenamefont {Absar}\ and\ \citenamefont
		{Coleman}(1976)}]{absar_one_1976}%
	\BibitemOpen
	\bibfield  {author} {\bibinfo {author} {\bibfnamefont {I.}~\bibnamefont
			{Absar}}\ and\ \bibinfo {author} {\bibfnamefont {A.}~\bibnamefont
			{Coleman}},\ }\bibfield  {title} {\bibinfo {title} {One electron orbitals
			intrinsic to the reduced hamiltonian},\ }\href
	{https://doi.org/10.1016/0009-2614(76)80342-9} {\bibfield  {journal}
		{\bibinfo  {journal} {Chemical Physics Letters}\ }\textbf {\bibinfo {volume}
			{39}},\ \bibinfo {pages} {609} (\bibinfo {year} {1976})}\BibitemShut
	{NoStop}%
	\bibitem [{\citenamefont {Harriman}(1978)}]{harriman_geometry_1978}%
	\BibitemOpen
	\bibfield  {author} {\bibinfo {author} {\bibfnamefont {J.~E.}\ \bibnamefont
			{Harriman}},\ }\bibfield  {title} {\bibinfo {title} {Geometry of density
			matrices. ii. reduced density matrices and n-representability},\ }\href
	{https://doi.org/10.1103/PhysRevA.17.1257} {\bibfield  {journal} {\bibinfo
			{journal} {Phys. Rev. A}\ }\textbf {\bibinfo {volume} {17}},\ \bibinfo
		{pages} {1257} (\bibinfo {year} {1978})}\BibitemShut {NoStop}%
	\bibitem [{\citenamefont {Mazziotti}(2002)}]{mazziotti_purification_2002}%
	\BibitemOpen
	\bibfield  {author} {\bibinfo {author} {\bibfnamefont {D.~A.}\ \bibnamefont
			{Mazziotti}},\ }\bibfield  {title} {\bibinfo {title} {Purification of
			correlated reduced density matrices},\ }\href
	{https://doi.org/10.1103/PhysRevE.65.026704} {\bibfield  {journal} {\bibinfo
			{journal} {Phys. Rev. E}\ }\textbf {\bibinfo {volume} {65}},\ \bibinfo
		{pages} {026704} (\bibinfo {year} {2002})}\BibitemShut {NoStop}%
	\bibitem [{\citenamefont {Colmenero}\ \emph {et~al.}(1993)\citenamefont
		{Colmenero}, \citenamefont {Pérez~del Valle},\ and\ \citenamefont
		{Valdemoro}}]{colmenero_approximating_1993}%
	\BibitemOpen
	\bibfield  {author} {\bibinfo {author} {\bibfnamefont {F.}~\bibnamefont
			{Colmenero}}, \bibinfo {author} {\bibfnamefont {C.}~\bibnamefont {P\'erez~del
				Valle}},\ and\ \bibinfo {author} {\bibfnamefont {C.}~\bibnamefont
			{Valdemoro}},\ }\bibfield  {title} {\bibinfo {title} {Approximating
			\textit{q}-order reduced density matrices in terms of the lower-order ones.
			{I}. {General} relations},\ }\href {https://doi.org/10.1103/PhysRevA.47.971}
	{\bibfield  {journal} {\bibinfo  {journal} {Physical Review A}\ }\textbf
		{\bibinfo {volume} {47}},\ \bibinfo {pages} {971} (\bibinfo {year}
		{1993})}\BibitemShut {NoStop}%
	\bibitem [{\citenamefont {Yasuda}\ and\ \citenamefont
		{Nakatsuji}(1997)}]{yasuda_direct_1997}%
	\BibitemOpen
	\bibfield  {author} {\bibinfo {author} {\bibfnamefont {K.}~\bibnamefont
			{Yasuda}}\ and\ \bibinfo {author} {\bibfnamefont {H.}~\bibnamefont
			{Nakatsuji}},\ }\bibfield  {title} {\bibinfo {title} {Direct determination of
			the quantum-mechanical density matrix using the density equation. {II}.},\
	}\href {https://doi.org/10.1103/PhysRevA.56.2648} {\bibfield  {journal}
		{\bibinfo  {journal} {Physical Review A}\ }\textbf {\bibinfo {volume} {56}},\
		\bibinfo {pages} {2648} (\bibinfo {year} {1997})}\BibitemShut {NoStop}%
	\bibitem [{\citenamefont {Joost}\ \emph {et~al.}(2022)\citenamefont {Joost},
		\citenamefont {Schl\"unzen}, \citenamefont {Ohldag}, \citenamefont {Bonitz},
		\citenamefont {Lackner},\ and\ \citenamefont
		{B\v rezinov\'a}}]{joost_dynamically_2022}%
	\BibitemOpen
	\bibfield  {author} {\bibinfo {author} {\bibfnamefont {J.-P.}\ \bibnamefont
			{Joost}}, \bibinfo {author} {\bibfnamefont {N.}~\bibnamefont {Schl\"unzen}},
		\bibinfo {author} {\bibfnamefont {H.}~\bibnamefont {Ohldag}}, \bibinfo
		{author} {\bibfnamefont {M.}~\bibnamefont {Bonitz}}, \bibinfo {author}
		{\bibfnamefont {F.}~\bibnamefont {Lackner}},\ and\ \bibinfo {author}
		{\bibfnamefont {I.}~\bibnamefont {B\v rezinov\'a}},\ }\bibfield  {title}
	{\bibinfo {title} {Dynamically screened ladder approximation: {Simultaneous}
			treatment of strong electronic correlations and dynamical screening out of
			equilibrium},\ }\href {https://doi.org/10.1103/PhysRevB.105.165155}
	{\bibfield  {journal} {\bibinfo  {journal} {Physical Review B}\ }\textbf
		{\bibinfo {volume} {105}},\ \bibinfo {pages} {165155} (\bibinfo {year}
		{2022})}\BibitemShut {NoStop}%
	\bibitem [{\citenamefont {Penrose}\ and\ \citenamefont
		{Onsager}(1956)}]{penrose_bose_1956}%
	\BibitemOpen
	\bibfield  {author} {\bibinfo {author} {\bibfnamefont {O.}~\bibnamefont
			{Penrose}}\ and\ \bibinfo {author} {\bibfnamefont {L.}~\bibnamefont
			{Onsager}},\ }\bibfield  {title} {\bibinfo {title} {Bose-einstein
			condensation and liquid helium},\ }\href
	{https://doi.org/10.1103/PhysRev.104.576} {\bibfield  {journal} {\bibinfo
			{journal} {Phys. Rev.}\ }\textbf {\bibinfo {volume} {104}},\ \bibinfo {pages}
		{576} (\bibinfo {year} {1956})}\BibitemShut {NoStop}%
	\bibitem [{\citenamefont {Coleman}\ and\ \citenamefont
		{Yukalov}(2000)}]{coleman_reduced_2000}%
	\BibitemOpen
	\bibfield  {author} {\bibinfo {author} {\bibfnamefont {A.~J.}\ \bibnamefont
			{Coleman}}\ and\ \bibinfo {author} {\bibfnamefont {V.~I.}\ \bibnamefont
			{Yukalov}},\ }\href@noop {} {\emph {\bibinfo {title} {Reduced Density
				Matrices, Coulson's Challenge}}}\ (\bibinfo  {publisher} {Springer-Verlag
		Berlin Heidelberg},\ \bibinfo {year} {2000})\BibitemShut {NoStop}%
	\bibitem [{\citenamefont {Gigena}\ \emph {et~al.}(2021)\citenamefont {Gigena},
		\citenamefont {Di~Tullio},\ and\ \citenamefont
		{Rossignoli}}]{gigena_many-body_2021}%
	\BibitemOpen
	\bibfield  {author} {\bibinfo {author} {\bibfnamefont {N.}~\bibnamefont
			{Gigena}}, \bibinfo {author} {\bibfnamefont {M.}~\bibnamefont {Di~Tullio}},\
		and\ \bibinfo {author} {\bibfnamefont {R.}~\bibnamefont {Rossignoli}},\
	}\bibfield  {title} {\bibinfo {title} {Many-body entanglement in fermion
			systems},\ }\href {https://doi.org/10.1103/PhysRevA.103.052424} {\bibfield
		{journal} {\bibinfo  {journal} {Physical Review A}\ }\textbf {\bibinfo
			{volume} {103}},\ \bibinfo {pages} {052424} (\bibinfo {year}
		{2021})}\BibitemShut {NoStop}%
	\bibitem [{\citenamefont {Ferreira}\ \emph {et~al.}(2022)\citenamefont
		{Ferreira}, \citenamefont {Maciel}, \citenamefont {Vianna},\ and\
		\citenamefont {Iemini}}]{ferreira_quantum_2022}%
	\BibitemOpen
	\bibfield  {author} {\bibinfo {author} {\bibfnamefont {D.~L.~B.}\
			\bibnamefont {Ferreira}}, \bibinfo {author} {\bibfnamefont {T.~O.}\
			\bibnamefont {Maciel}}, \bibinfo {author} {\bibfnamefont {R.~O.}\
			\bibnamefont {Vianna}},\ and\ \bibinfo {author} {\bibfnamefont
			{F.}~\bibnamefont {Iemini}},\ }\bibfield  {title} {\bibinfo {title} {Quantum
			correlations, entanglement spectrum, and coherence of the two-particle
			reduced density matrix in the extended {Hubbard} model},\ }\href
	{https://doi.org/10.1103/PhysRevB.105.115145} {\bibfield  {journal} {\bibinfo
			{journal} {Physical Review B}\ }\textbf {\bibinfo {volume} {105}},\ \bibinfo
		{pages} {115145} (\bibinfo {year} {2022})}\BibitemShut {NoStop}%
	\bibitem [{\citenamefont {Cheuk}\ \emph {et~al.}(2015)\citenamefont {Cheuk},
		\citenamefont {Nichols}, \citenamefont {Okan}, \citenamefont {Gersdorf},
		\citenamefont {Ramasesh}, \citenamefont {Bakr}, \citenamefont {Lompe},\ and\
		\citenamefont {Zwierlein}}]{cheuk_quantum-gas_2015}%
	\BibitemOpen
	\bibfield  {author} {\bibinfo {author} {\bibfnamefont {L.~W.}\ \bibnamefont
			{Cheuk}}, \bibinfo {author} {\bibfnamefont {M.~A.}\ \bibnamefont {Nichols}},
		\bibinfo {author} {\bibfnamefont {M.}~\bibnamefont {Okan}}, \bibinfo {author}
		{\bibfnamefont {T.}~\bibnamefont {Gersdorf}}, \bibinfo {author}
		{\bibfnamefont {V.~V.}\ \bibnamefont {Ramasesh}}, \bibinfo {author}
		{\bibfnamefont {W.~S.}\ \bibnamefont {Bakr}}, \bibinfo {author}
		{\bibfnamefont {T.}~\bibnamefont {Lompe}},\ and\ \bibinfo {author}
		{\bibfnamefont {M.~W.}\ \bibnamefont {Zwierlein}},\ }\bibfield  {title}
	{\bibinfo {title} {Quantum-{Gas} {Microscope} for {Fermionic} {Atoms}},\
	}\href {https://doi.org/10.1103/PhysRevLett.114.193001} {\bibfield  {journal}
		{\bibinfo  {journal} {Physical Review Letters}\ }\textbf {\bibinfo {volume}
			{114}},\ \bibinfo {pages} {193001} (\bibinfo {year} {2015})}\BibitemShut
	{NoStop}%
	\bibitem [{\citenamefont {Parsons}\ \emph {et~al.}(2015)\citenamefont
		{Parsons}, \citenamefont {Huber}, \citenamefont {Mazurenko}, \citenamefont
		{Chiu}, \citenamefont {Setiawan}, \citenamefont {Wooley-Brown}, \citenamefont
		{Blatt},\ and\ \citenamefont {Greiner}}]{parsons_site-resolved_2015}%
	\BibitemOpen
	\bibfield  {author} {\bibinfo {author} {\bibfnamefont {M.~F.}\ \bibnamefont
			{Parsons}}, \bibinfo {author} {\bibfnamefont {F.}~\bibnamefont {Huber}},
		\bibinfo {author} {\bibfnamefont {A.}~\bibnamefont {Mazurenko}}, \bibinfo
		{author} {\bibfnamefont {C.~S.}\ \bibnamefont {Chiu}}, \bibinfo {author}
		{\bibfnamefont {W.}~\bibnamefont {Setiawan}}, \bibinfo {author}
		{\bibfnamefont {K.}~\bibnamefont {Wooley-Brown}}, \bibinfo {author}
		{\bibfnamefont {S.}~\bibnamefont {Blatt}},\ and\ \bibinfo {author}
		{\bibfnamefont {M.}~\bibnamefont {Greiner}},\ }\bibfield  {title} {\bibinfo
		{title} {Site-{Resolved} {Imaging} of {Fermionic} {Li} 6 in an {Optical}
			{Lattice}},\ }\href {https://doi.org/10.1103/PhysRevLett.114.213002}
	{\bibfield  {journal} {\bibinfo  {journal} {Physical Review Letters}\
		}\textbf {\bibinfo {volume} {114}},\ \bibinfo {pages} {213002} (\bibinfo
		{year} {2015})}\BibitemShut {NoStop}%
	\bibitem [{\citenamefont {Haller}\ \emph {et~al.}(2015)\citenamefont {Haller},
		\citenamefont {Hudson}, \citenamefont {Kelly}, \citenamefont {Cotta},
		\citenamefont {Peaudecerf}, \citenamefont {Bruce},\ and\ \citenamefont
		{Kuhr}}]{haller_single-atom_2015}%
	\BibitemOpen
	\bibfield  {author} {\bibinfo {author} {\bibfnamefont {E.}~\bibnamefont
			{Haller}}, \bibinfo {author} {\bibfnamefont {J.}~\bibnamefont {Hudson}},
		\bibinfo {author} {\bibfnamefont {A.}~\bibnamefont {Kelly}}, \bibinfo
		{author} {\bibfnamefont {D.~A.}\ \bibnamefont {Cotta}}, \bibinfo {author}
		{\bibfnamefont {B.}~\bibnamefont {Peaudecerf}}, \bibinfo {author}
		{\bibfnamefont {G.~D.}\ \bibnamefont {Bruce}},\ and\ \bibinfo {author}
		{\bibfnamefont {S.}~\bibnamefont {Kuhr}},\ }\bibfield  {title} {\bibinfo
		{title} {Single-atom imaging of fermions in a quantum-gas microscope},\
	}\href {https://doi.org/10.1038/nphys3403} {\bibfield  {journal} {\bibinfo
			{journal} {Nature Physics}\ }\textbf {\bibinfo {volume} {11}},\ \bibinfo
		{pages} {738} (\bibinfo {year} {2015})}\BibitemShut {NoStop}%
	\bibitem [{\citenamefont {Edge}\ \emph {et~al.}(2015)\citenamefont {Edge},
		\citenamefont {Anderson}, \citenamefont {Jervis}, \citenamefont {McKay},
		\citenamefont {Day}, \citenamefont {Trotzky},\ and\ \citenamefont
		{Thywissen}}]{edge_imaging_2015}%
	\BibitemOpen
	\bibfield  {author} {\bibinfo {author} {\bibfnamefont {G.~J.~A.}\
			\bibnamefont {Edge}}, \bibinfo {author} {\bibfnamefont {R.}~\bibnamefont
			{Anderson}}, \bibinfo {author} {\bibfnamefont {D.}~\bibnamefont {Jervis}},
		\bibinfo {author} {\bibfnamefont {D.~C.}\ \bibnamefont {McKay}}, \bibinfo
		{author} {\bibfnamefont {R.}~\bibnamefont {Day}}, \bibinfo {author}
		{\bibfnamefont {S.}~\bibnamefont {Trotzky}},\ and\ \bibinfo {author}
		{\bibfnamefont {J.~H.}\ \bibnamefont {Thywissen}},\ }\bibfield  {title}
	{\bibinfo {title} {Imaging and addressing of individual fermionic atoms in an
			optical lattice},\ }\href {https://doi.org/10.1103/PhysRevA.92.063406}
	{\bibfield  {journal} {\bibinfo  {journal} {Physical Review A}\ }\textbf
		{\bibinfo {volume} {92}},\ \bibinfo {pages} {063406} (\bibinfo {year}
		{2015})}\BibitemShut {NoStop}%
	\bibitem [{\citenamefont {Omran}\ \emph {et~al.}(2015)\citenamefont {Omran},
		\citenamefont {Boll}, \citenamefont {Hilker}, \citenamefont {Kleinlein},
		\citenamefont {Salomon}, \citenamefont {Bloch},\ and\ \citenamefont
		{Gross}}]{omran_microscopic_2015}%
	\BibitemOpen
	\bibfield  {author} {\bibinfo {author} {\bibfnamefont {A.}~\bibnamefont
			{Omran}}, \bibinfo {author} {\bibfnamefont {M.}~\bibnamefont {Boll}},
		\bibinfo {author} {\bibfnamefont {T.~A.}\ \bibnamefont {Hilker}}, \bibinfo
		{author} {\bibfnamefont {K.}~\bibnamefont {Kleinlein}}, \bibinfo {author}
		{\bibfnamefont {G.}~\bibnamefont {Salomon}}, \bibinfo {author} {\bibfnamefont
			{I.}~\bibnamefont {Bloch}},\ and\ \bibinfo {author} {\bibfnamefont
			{C.}~\bibnamefont {Gross}},\ }\bibfield  {title} {\bibinfo {title}
		{Microscopic {Observation} of {Pauli} {Blocking} in {Degenerate} {Fermionic}
			{Lattice} {Gases}},\ }\href {https://doi.org/10.1103/PhysRevLett.115.263001}
	{\bibfield  {journal} {\bibinfo  {journal} {Physical Review Letters}\
		}\textbf {\bibinfo {volume} {115}},\ \bibinfo {pages} {263001} (\bibinfo
		{year} {2015})}\BibitemShut {NoStop}%
	\bibitem [{\citenamefont {Cai}\ \emph {et~al.}(2022)\citenamefont {Cai},
		\citenamefont {Allman}, \citenamefont {Sabharwal},\ and\ \citenamefont
		{Wright}}]{cai_persistent_2022}%
	\BibitemOpen
	\bibfield  {author} {\bibinfo {author} {\bibfnamefont {Y.}~\bibnamefont
			{Cai}}, \bibinfo {author} {\bibfnamefont {D.~G.}\ \bibnamefont {Allman}},
		\bibinfo {author} {\bibfnamefont {P.}~\bibnamefont {Sabharwal}},\ and\
		\bibinfo {author} {\bibfnamefont {K.~C.}\ \bibnamefont {Wright}},\ }\bibfield
	{title} {\bibinfo {title} {Persistent {Currents} in {Rings} of {Ultracold}
			{Fermionic} {Atoms}},\ }\href
	{https://doi.org/10.1103/PhysRevLett.128.150401} {\bibfield  {journal}
		{\bibinfo  {journal} {Physical Review Letters}\ }\textbf {\bibinfo {volume}
			{128}},\ \bibinfo {pages} {150401} (\bibinfo {year} {2022})}\BibitemShut
	{NoStop}%
\end{thebibliography}
%

\end{document}